\documentclass[%
reprint,
superscriptaddress,
nofootinbib,
amsmath,amssymb,
aps,
prb,
floatfix,
longbibliography
]{revtex4-2}

\usepackage{graphicx}
\usepackage{dcolumn}
\usepackage{bm}
\usepackage[colorlinks=true, linkcolor=blue,urlcolor=blue,citecolor=blue]{hyperref}
\usepackage[mathlines]{lineno}

\usepackage[ignoreunlbld,norefs,nocites]{refcheck}

\usepackage{ulem}

\begin{document}

\preprint{APS/123-QED}

\title{Ground-state properties of electron-electron biwire systems}

\author{Rajesh O.\ Sharma}
\email{sharmarajesh0387@gmail.com}
\affiliation{%
Department of Physics, Panjab University, Chandigarh-160014, India}%

\author{N.\ D.\ Drummond}
\affiliation{Department of Physics, Lancaster University, Lancaster LA1 4YB, United Kingdom}%

\author{Vinod Ashokan}
\affiliation{Department of Physics, Dr.\ B.\ R.\ Ambedkar National Institute of Technology, Jalandhar (Punjab) 144011, India}

\author{K.\ N.\ Pathak}%
\affiliation{%
Department of Physics, Panjab University, Chandigarh-160014, India}%

\author{Klaus Morawetz}
\affiliation{M\"{u}nster University of Applied Sciences, Stegerwaldstrasse 39, 48565 Steinfurt, Germany}%
\affiliation{International Institute of Physics - UFRN, Campus Universit\'ario Lagoa nova, 59078-970 Natal, Brazil}%

\date{\today}

\begin{abstract}
The correlation between electrons in different quantum wires is expected to affect the electronic properties of quantum electron-electron biwire systems. Here, we use the variational Monte Carlo method to study the ground-state properties of parallel, infinitely thin electron-electron biwires for several electron densities ($r_\text{s}$) and interwire separations ($d$). Specifically, the ground-state energy, the correlation energy, the interaction energy, the pair-correlation function (PCF), the static structure factor (SSF), and the momentum distribution (MD) function are calculated. We find that the interaction energy increases as $\ln(d)$ for $d\to 0$ and it decreases as $d^{-2}$ when $d\to \infty$. The PCF shows oscillatory behavior at all densities considered here. As two parallel wires approach each other, interwire correlations increase while intrawire correlations decrease as evidenced by the behavior of the PCF, SSF, and MD\@. The system evolves  from two monowires of density parameter $r_\text{s}$ to a single monowire of density parameter $r_\text{s}/2$ as $d$ is reduced from infinity to zero. The MD reveals Tomonaga-Luttinger (TL) liquid behavior with a power-law nature near $k_\text{F}$ even in the presence of an extra interwire interaction between the electrons in biwire systems. It is observed that when $d$ is reduced the MD decreases for $k<k_\text{F}$ and increases for $k>k_\text{F}$, similar to its behavior with increasing $r_\text{s}$.  The TL liquid exponent is extracted by fitting the MD data near $k_\text{F}$, from which the TL liquid interaction parameter $K_{\rho}$ is calculated. The value of the TL parameter is found to be in agreement with that of a single wire for large separation between the two wires.
\end{abstract}

\maketitle


\section{Introduction}
One-dimensional (1D) systems of interacting fermions have gained considerable interest in both experimental \cite{Goni1991,
Altmann2001,
Nagao2006,
Hong2016} and theoretical \cite{Friesen1980, DasSarma1985,
Schulz1993, Tanatar1998c, Moudgil2010a} fields due to their wide range of interesting quantum properties and potential applications in various areas of electronics, sensors, and medicine. The simplest theoretical model of interacting electrons is the homogeneous electron gas, in which electrons are neutralized by a uniform, positively charged background. Fermi liquid theory, which works very well for interacting fermions in two- and three-dimensional systems, breaks down in 1D systems of fermions. The Tomonaga-Luttinger (TL) liquid is a standard model for describing the physical properties of 1D electron systems \cite{Tomonaga1950, Luttinger1963, Haldane1982, Giamarchi2003}. There have been extensive theoretical and computational studies of electron correlation effects in isolated 1D interacting systems using various techniques such as the random phase approximation (RPA) \cite{Bala2014, Ashokan2018, Morawetz2018, Ashokan2020}, Singwi, Tosi, Land, and Sj\"{o}lander (STLS) \cite{Tanatar1999, Demirel1999a, Garg2008, Sharma2018c}, and quantum Monte Carlo (QMC) methods \cite{Casula2006, Shulenburger2008, Lee2011b, Ashokan2018a}.

Two-dimensional systems of coupled, parallel quantum layers (electron-electron or electron-hole bilayers) show many unique phenomena \cite{Senatore2003, Kou2014, Sharma2016, Butov2017, Sharma2017, LopezRios2018, Sharma2018}. Similarly, in 1D systems the additional interaction between charge carriers residing in different wires yields quantum properties such as non-Abelian topological phases (edge properties) \cite{Yang2020, Li2020, Meng2019, Fuji2019, Iadecola2019}, Coulomb drag between wires \cite{Zhou2019, Debray2002, Tanatar1998b}, nonadditive dispersion \cite{Misquitta2014, Drummond2007, Dobson2006, Chang1971}, enhancement in the onset of Wigner crystallization \cite{Moudgil2010}, and formation of biexcitons \cite{Zhang2008, Szafran2005, Tsuchiya2001}. Because of these interesting properties, coupled parallel quantum wires have gained significant attention in the research community.

The majority of theoretical work on 1D biwire systems is based on the RPA \cite{Gold1992} and STLS \cite{Moudgil2010, Saini2004, Moudgil2000, Mutluay1997a, Thakur1997, Wang1995a} methods. Although these methods have been used to perform elaborate calculations of various ground-state properties of biwire systems, their findings have remained unverified until recently due to the unavailability of simulation data and experimental results. Drummond and Needs \cite{Drummond2007} used QMC methods to obtain the binding energy of coupled metallic wires. However, there are further interesting properties to be studied.

In this paper we use the variational Monte Carlo (VMC) method to investigate inter- and intra-wire correlation effects on the ground-state properties of electron-electron biwire (EEBW) systems. Simulation results obtained with QMC methods such as VMC and the more accurate diffusion Monte Carlo (DMC) method can be treated as benchmarks in the absence of experimental results. In fact, for benchmarking theory, QMC may be even better suited than experiments, in that it provides an essentially exact solution to a well-defined model without effects such as disorder and vibrations that inevitably complicate the interpretation of experimental data. In 1D, fixed-node DMC is an exact fermion ground-state method because the nodal surface is known exactly. However, DMC is much more computationally costly than VMC\@. Since VMC is able to extract most of the correlation energy of 1D electron systems \cite{Lee2011b}, it is sufficiently accurate in this case. We report VMC results for the momentum distribution (MD) functions, energies, pair-correlation functions (PCFs), and static structure factors (SSFs) of infinitely thin quantum biwires at a variety of densities and interwire separations. The PCF and SSF provide useful information about electronic correlations and are useful quantities for assessing the nature of the ground state, while the MD is used to extract parameters relating to the TL liquid properties of the system. One of the motivations of the present work is to see the effects of interwire interactions on the TL parameters. The results for the MD in particular show the non-Fermi-liquid character of the system. The total-energy data that we provide may be regarded as benchmarks for future theoretical work.

The paper is structured as follows: We describe the EEBW model in Sec.\ \ref{sec:II}. In Sec.\ \ref{sec:III} we outline the VMC method and provide the details of our approach. In Sec.\ \ref{sec:IV} we report the ground-state energy, PCF, SSF, and MD of an infinitely thin EEBW system. We discuss the effects of finite system sizes on the various observables mentioned above in Sec.\ \ref{sec:IV-FSE}. Finally, the conclusions are in Sec.\ \ref{sec:V}.

\begin{figure}[!htbp]
\centering
\includegraphics[width=.48\textwidth]{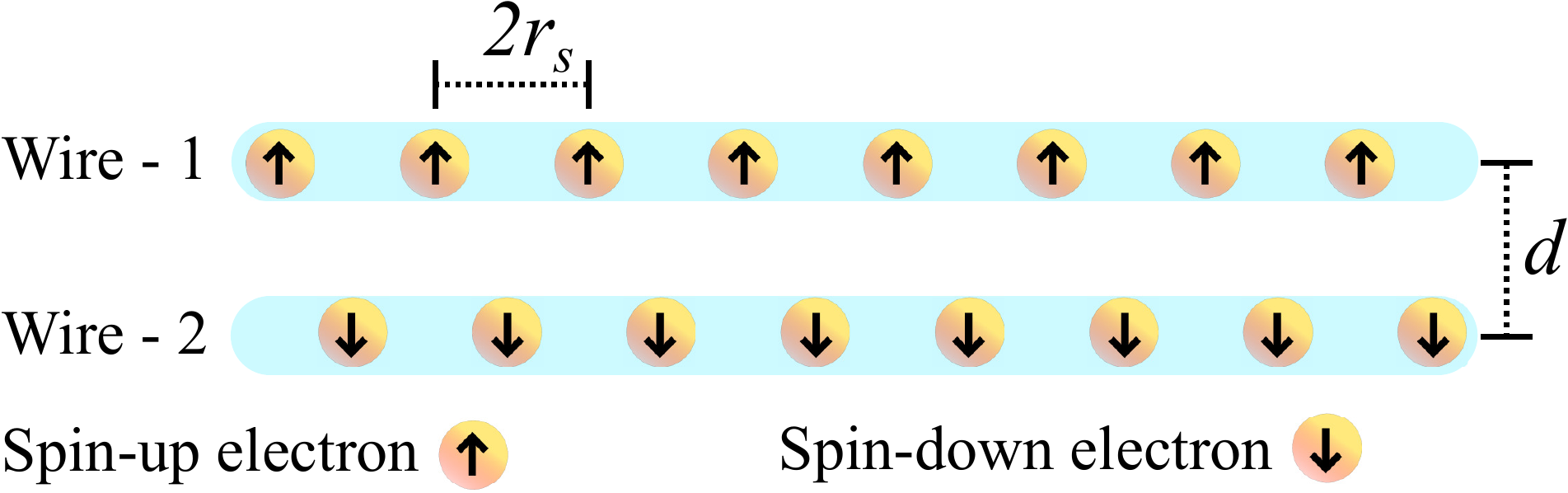}
\caption{\label{fig:model} Cartoon representation of the EEBW model. Both wires are identical in all aspects except the spin of the electrons. The mean distance between the electrons in each wire is $2r_\text{s}$.}
\end{figure}

\section{Biwire model}\label{sec:II}
We consider an EEBW system consisting of two parallel, infinitely thin quantum wires that are separated by a distance $d$ as shown in Fig.\ \ref{fig:model}.  The top wire contains only spin-up electrons while the bottom wire contains only spin-down electrons. We assume that the electrons in each wire are embedded in a uniform, positive background to maintain charge neutrality. Most of the experimental studies of biwire systems \cite{Hansen1987, Demel1988, Debray2001} have used identical wires; therefore we focus on identical EEBWs. The electron masses and the electron densities are chosen to be the same in the two wires. The electron density $n$ in each wire is determined by the dimensionless density parameter $r_\text{s}=1/(2na_\text{B})$, where $a_\text{B}=\epsilon/(e^2m_\text{e})$ is the Bohr radius, $\epsilon$ is the background dielectric constant, and $m_\text{e}$ is the electron effective mass. We use effective Hartree atomic units ($\hbar =|e|= m_\text{e} = 4\pi\epsilon = 1$) throughout the remainder of this article. %

The interaction potential between an isolated pair of electrons in the same wire is $1/|x|$, and the potential between isolated electrons in opposite wires is $1/\sqrt{x^2+d^2}$, where $x$ is the component of electron separation in the direction of the wires. We write the Hamiltonian of the infinite EEBW system with $N$ electrons per wire as
\begin{align}
\hat{H}=&
-\dfrac{1}{2}\sum_{i=1}^N\left( \dfrac{\partial^2}{\partial x_{i,1}^2}+\dfrac{\partial^2}{\partial x_{i,2}^2}\right) \nonumber \\
&+\sum_{i<j} \left[ V(|x_{i,1} - x_{j,1}|,0)+V(|x_{i,2} - x_{j,2}|,0)\right] \nonumber \\
&+\sum_{i,j} V(|x_{i,1}-x_{j,2}|,d)
+ NV_{\textrm{Mad}},
\end{align}
where $x_{i,m}$ is the position of electron $i$ in wire $m$, $V(x,z)$ is the 1D Ewald interaction between electrons with in-wire separation $x$ and out-of-wire separation $z$, and $V_{\textrm{Mad}}$ is the Madelung constant \cite{Saunders1994}. It is known that the ground-state many-body wave function of a system of fermions interacting via the Coulomb interaction in an infinitely thin 1D wire has nodes at all coalescence points, irrespective of the orientation of the spins \cite{Lee2011b}. Therefore, the paramagnetic and ferromagnetic states are degenerate and the Lieb-Mattis theorem \cite{Lieb1962} does not apply. As a result, the ground-state energy only depends on the density rather than on the spin polarization. For computational convenience, we consider both wires to be fully spin-polarized in our EEBW model.

\section{Method}\label{sec:III}
In this section, we present our VMC method and the parameters associated with it. We describe the trial wave functions and the form of the single-particle orbitals used to calculate the ground-state energy of the EEBW system. We have used the {\sc casino} \cite{Needs2020} code to perform VMC calculations.

\subsection{Variational Monte Carlo} 
In the VMC technique, the expectation value of the Hamiltonian $\hat{H}$ with respect to a trial wave function $\Psi_\text{T}$ is calculated using importance-sampled Monte Carlo integration \cite{Foulkes2001}. The trial wave function contains a number of variable parameters whose values are optimized by the use of variational principles. VMC provides an upper bound on the exact ground-state energy. The variational energy expectation value of $\hat{H}$ with trial wave function $\Psi_\text{T}$ is given by
\begin{equation}
\begin{split}
\langle E_\text{T}\rangle &=
\dfrac{\int \Psi^*_\text{T}(\mathbf{X})\hat{H}\Psi_\text{T}(\mathbf{X}) \, d\mathbf{X}}{\int\Psi^*_\text{T}(\mathbf{X})\Psi_\text{T}(\mathbf{X}) \, d\mathbf{X}}\\
& =\int\dfrac{|\Psi_\text{T}(\mathbf{X}')|^2}{\int|\Psi_\text{T}(\mathbf{X}')|^2 \, d\mathbf{X}'} E_\text{L}(\mathbf{X}) \, d\mathbf{X},
\end{split}
\end{equation}%
where $\mathbf{X}$ is a vector of all electron $x$ coordinates and
 $E_\text{L}(\mathbf{X})=\Psi^{-1}_\text{T}(\mathbf{X})\hat{H}\Psi_\text{T}(\mathbf{X})$ is the local energy.

\subsection{Trial wave functions} 

Our many-body trial wave function is of Slater-Jastrow-backflow type and consists of Slater determinants of plane-wave orbitals multiplied by a Jastrow correlation factor. The Jastrow factor contains polynomial and plane-wave expansions in electron-electron separation.  We consider electrons in different wires to be distinguishable; therefore the trial wave function for a biwire consists of a product of two Slater determinants. The Slater-Jastrow trial wave function is
\begin{equation}
\Psi_\text{T}=
D(\phi^{\uparrow}(x))
D(\phi^{\downarrow}(x))
e^{J(\mathbf{X})},
\end{equation}
where $\phi^{\uparrow}$ represents orbitals for spin-up electron, $D$ is the Slater determinant, and $e^{J(\mathbf{X})}$ is a Jastrow factor, which describes the correlations between the charge carriers within the wire and between the wires. Plane-wave orbitals
\begin{equation}
\phi(x)=\exp(ikx)
\end{equation}
with wavenumbers up to $k_\text{F} = \pi/(2r_\text{s} )$ were used in the Slater determinants. We look at systems with time-reversal symmetry, so that the wave function $\Psi_\text{T}(x)$ is real.

We use a backflow transformation \cite{LopezRios2006a}. In this technique, coordinates of electrons in the Slater determinants are replaced by ``quasiparticle coordinates'' related to the actual electron positions by backflow functions consisting of polynomial expansions in the electron $x$ separation up to 8th order \cite{LopezRios2006a}. We use separate terms for intra- and inter-wire electron pairs. Normally, backflow functions are used to improve the nodal surfaces of Slater determinants in VMC trial wave functions. For infinitely thin wires, Lee and Drummond \cite{Lee2011b} concluded that the divergence in the interaction potential at coalescence points at which the wave function does not vanish cannot be cancelled by a divergence in the kinetic energy, and hence the trial wave function must possess nodes at all of the coalescence points. Therefore, for this system the backflow transformation does not change the nodal surface, which is already exact, although it provides a compact parameterization of three-body correlations \cite{Lee2011b}.

We use {\sc casino}'s Jastrow factor \cite{Drummond2004a}, with a two-body polynomial $u$ term and a plane-wave term $p$. The $u$ term consists of an expansion in powers of electron-electron $x$ separation up to 8th order. The $p$ term is a Fourier expansion with 20 independent reciprocal-lattice points. These functions in the Jastrow factor and backflow function contain the free parameters which are optimized within the VMC method. We use non-reweighted variance minimization \cite{Umrigar1988, Drummond2005} followed by energy minimization \cite{Umrigar2007} to optimize the free parameters of the trial wave function. To optimize these parameters, we use $5\times 10^6$ statistically independent steps and 1024 configurations.

The VMC method is capable of giving highly accurate results for 1D systems. For example, Lee and Drummond \cite{Lee2011b} showed that a two-body Jastrow factor with backflow transformations can retrieve 99.9989(9)\% of the correlation energy within the VMC method for an infinitely thin wire at $r_\text{s} = 15$ and $N = 15$. For some representative cases we have checked that our VMC calculations agree with DMC results (see Sec.\ \ref{sec:V_DMC}). However, the ground-state energy and other observables are subject to finite-size effects due to the limited size of the simulation cell. Lee and Drummond have demonstrated that twist averaging \cite{Lin2001}, which has been shown to greatly reduce single-particle finite-size effects in two and three dimensions, is of limited use in 1D systems because momentum-quantization errors are systematic rather than quasirandom in 1D\@. In our work, ground-state energies are extrapolated to the thermodynamic limit to eliminate the finite-size bias. Finite-size effects appear to be negligibly small in the PCF, SSF, and MD for the largest system size considered in this paper (see Sec.\ \ref{sec:IV-FSE}).

\section{Results and discussion}\label{sec:IV}
For our VMC calculations of the energy, PCF, SSF, and MD, we consider an EEBW in a simulation cell of length $L = 2Nr_\text{s}$ subject to periodic boundary conditions, where $N$ is the number of electrons per wire. $N = 61$ was the largest system considered, for which the biwire system has $122$ electrons. To extrapolate the VMC energy to the thermodynamic limit, we also performed calculations with $N = 21$ and 41.

\begin{figure}[!htbp]
\centering
\includegraphics[width=.48\textwidth]{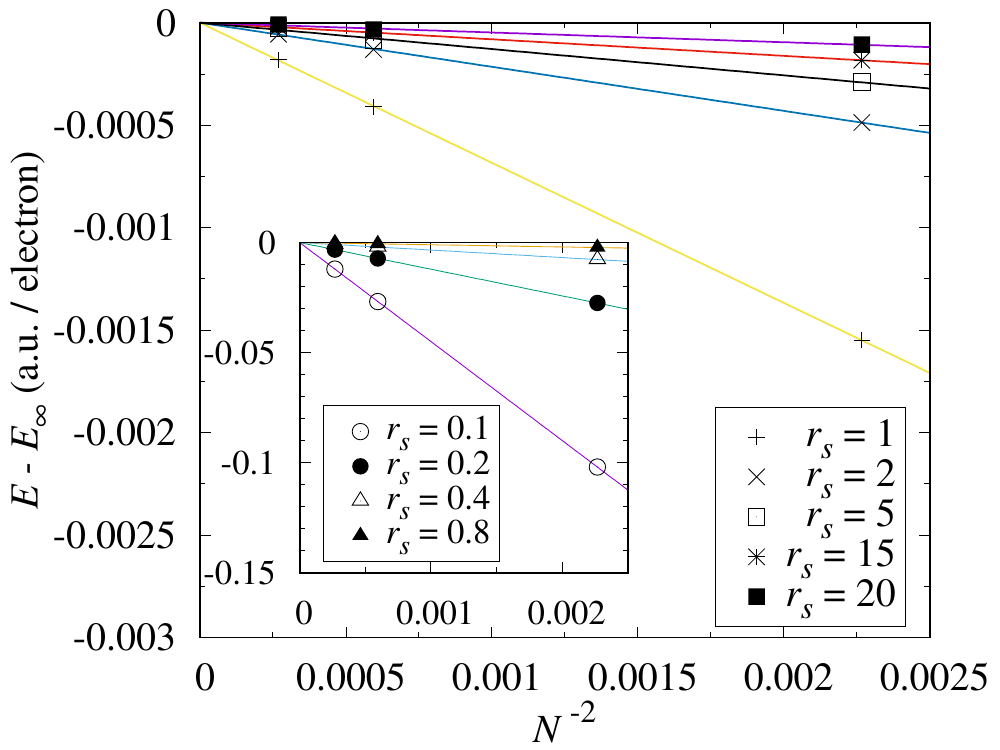}
\caption{\label{fig:e-en} EEBW VMC energies (a.u./electron) offset by the extrapolated $E_\infty$ vs.\ reciprocal of the square of the system size at interwire separation $d = 1$ a.u. Equation (\ref{eqn:Ethrmdnm}) is linearly fitted to the VMC energy data for different system sizes $N$ to obtain the asymptotic value of the ground-state energy per electron $E_\infty$.  Error bars are smaller than the size of the symbols.}
\end{figure}

\begin{figure}[!htbp]
\centering
\includegraphics[width=.48\textwidth]{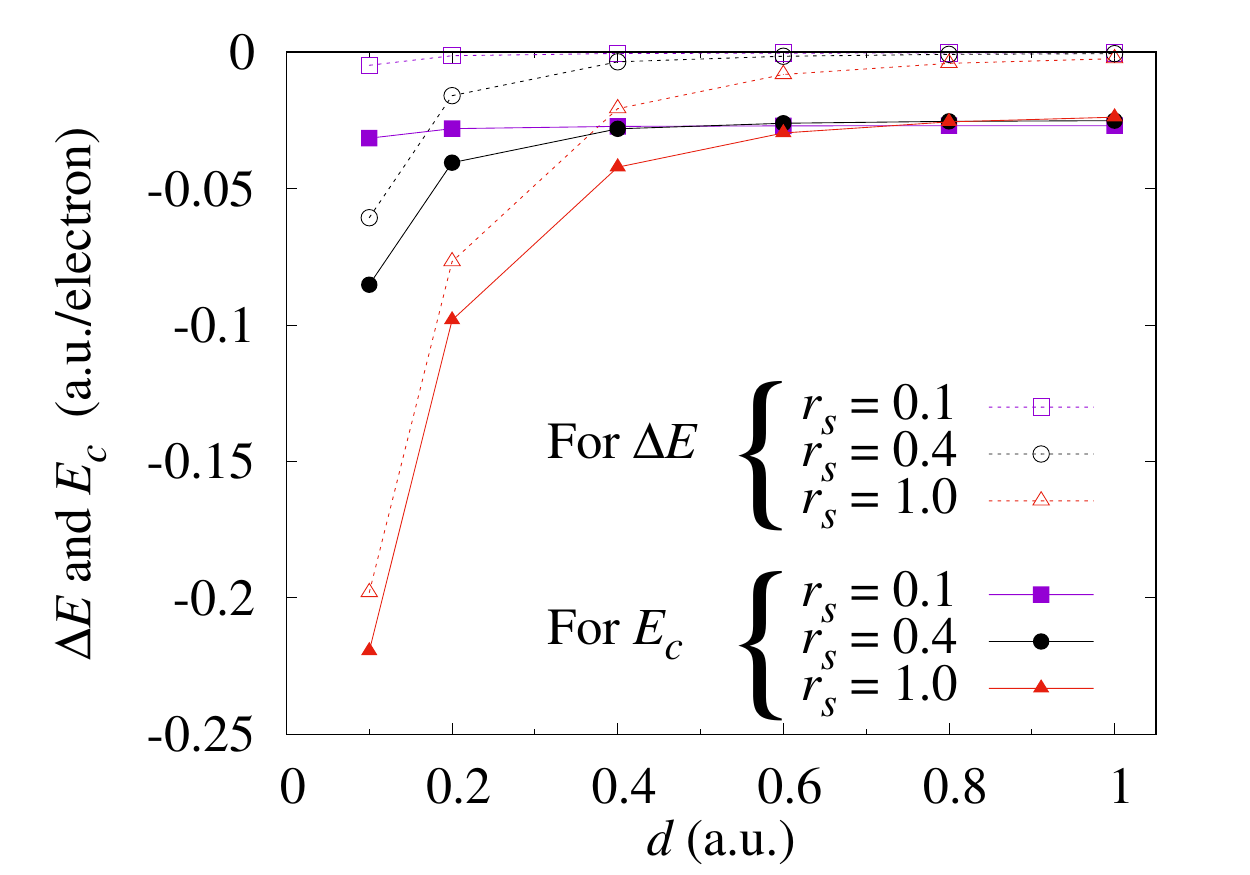}
\caption{\label{fig:BECE} Interaction energy $\Delta E$ (a.u./electron) and correlation energy $E_\text{c}$ (a.u./electron) are plotted as functions of interwire spacing $d$ for $r_\text{s} \leq 1$. Open symbols with dashed lines represent $\Delta E$ and closed symbols with solid lines are for $E_\text{c}$. $\Delta E$ and $E_\text{c}$ are calculated using $E_{\infty}$ for the wire and biwire systems.}
\end{figure}

\begin{figure}[!htbp]
\centering
\includegraphics[width=.48\textwidth]{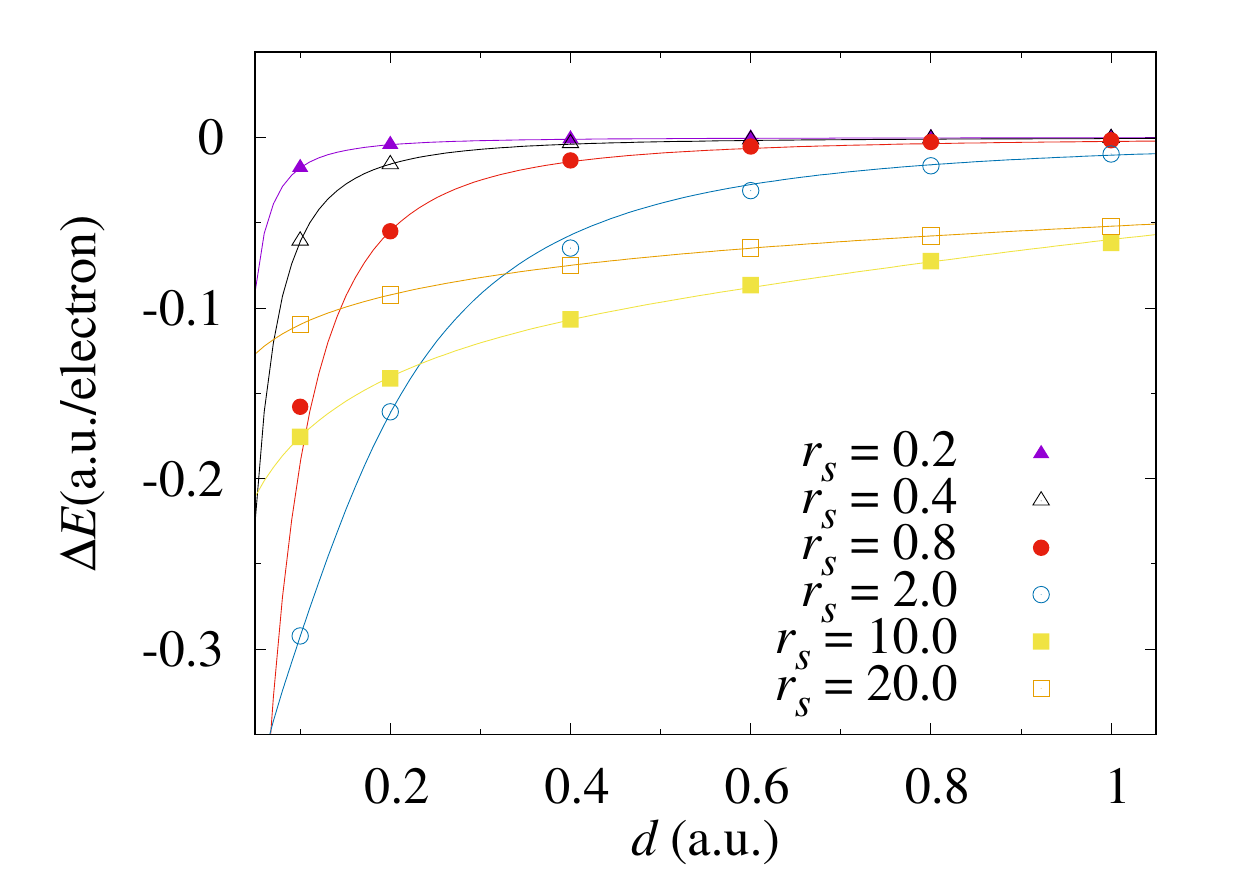}
\caption{\label{fig:BE-ndd}
Interwire interaction energy $\Delta E$ vs.\ $d$. Symbols represent our calculated data and solid lines show data obtained by fitting Eq.\ (\ref{eqn:BE-expC2}).
}
\end{figure}

\subsection{Energies} 
For the EEBW system we have calculated the VMC ground-state energy per electron for $r_\text{s} = 0.1$, 0.2, 0.4, 0.8, 1, 2, 5, 10, 15, and 20, and have reduced the interwire separation $d$ from $1$ to $0.1$ a.u.\ at an interval of $0.2$ a.u.\ for each $r_\text{s}$. It has been shown \cite{Lee2011b, Misquitta2014} that the total energy per electron for the 1D homogeneous electron gas scales with system size as
\begin{equation}
E(N) = E_{\infty} +\dfrac{B}{N^{2}},
\label{eqn:Ethrmdnm}
\end{equation}
where $E_{\infty}$ and $B$ are fitting parameters for any given $d$. Therefore, we have extrapolated the VMC energy per electron of the EEBW system to the thermodynamic limit using Eq.\ (\ref{eqn:Ethrmdnm}). Figure \ref{fig:e-en} shows that Eq.\ (\ref{eqn:Ethrmdnm}) fits our energy data well. These energies, calculated at various values of $r_\text{s}$ and $d$ for an EEBW, are tabulated in Table \ref{tab:table1} of Appendix \ref{ap-i}. In Appendix \ref{ap:chi2} we investigate finite size extrapolation using the formula proposed in Ref.\ \onlinecite{Shulenburger2009}. However, the resulting ground-state energies are almost the same.

We have calculated the correlation energy per electron $E_\text{c}$ and interaction energy per electron $\Delta E$ from the extrapolated ground-state energy ($E_{\infty}$), which are also included in Table \ref{tab:table1}. The total energy per electron of a biwire system is given by
\begin{equation}
E_{\rm bi}(d)=\left[2 e_{\rm mono} + \Delta e(d)\right] / (2 N) = E_{\rm mono} + \Delta E(d).
\end{equation}
Here, each lowercase $e$ represents a total energy and each uppercase $E$ represents an energy per electron. Note that the biwire system has a total of $2N$ electrons ($N$ on each wire). The interaction energy per electron is then given by
\begin{equation}
\Delta E(d) = E_\text{bi}(d) - E_\text{mono},
\label{eqn:BE}
\end{equation}
and the correlation energy per electron as
\begin{equation}
E_\text{c}(d) = E_\text{bi}(d) - E_\text{HF}.
\end{equation}
Here, $E_\text{mono}$, $E_\text{bi}$, and $E_\text{HF}$ are the ground-state energy per electron of a single wire, a biwire, and the Hartree-Fock (HF) energy per electron, respectively. Misquitta \textit{et al.}\ \cite{Misquitta2014} reported that the interaction energies at a given $d$ decay more slowly with system size. They extrapolated $\Delta E$ to the thermodynamic limit using equation
\begin{equation}
\Delta E(N) = \Delta E_{\infty} +\dfrac{B'}{N}.
\label{eqn:BEthrmdnm}
\end{equation}
We fitted our data with both Eqs.\ (\ref{eqn:Ethrmdnm}) and (\ref{eqn:BEthrmdnm}), and found that our interaction energy data are better described by Eq.\ (\ref{eqn:Ethrmdnm}). The reason for the better fitting is argued by Drummond and Needs \cite{Drummond2007} that when the difference of energies is taken out, most of the bias is canceled. The interaction energy and correlation energies shown in Table \ref{tab:table1} were calculated from the $E_{\infty}$ values obtained using Eq.\ (\ref{eqn:Ethrmdnm}).

Figure \ref{fig:BECE} shows $\Delta E$ and $E_\text{c}$ as functions of separation between two wires for high electron densities. The correlation energy per electron of the biwire $E_\text{c}(d)$ is the sum of the correlation energy per electron of the isolated single wire $E_{\rm c}^\text{mono}$ and the interaction energy per electron $\Delta E(d)$, i.e., $E_\text{c}(d) = E_{\rm c}^\text{mono} + \Delta E(d)$. Therefore, the dependence of correlation energy $E_\text{c}(d)$ on the wire separation $d$ is similar to $\Delta E(d)$. The interaction energy of the positive backgrounds of the two wires is $\ln(d^2)/(4r_\text{s})$~\cite{Saunders1994}. This suggests that at a given $r_\text{s}$, the interaction energy of a biwire may be represented by
\begin{equation}
\Delta E(d) = \frac{E_\text{mono}(r_\text{s}/2)-E_\text{mono}(r_\text{s})+\ln(d^2)/(4r_\text{s}) + Ad^2}{1+Bd^2+Cd^4},
\label{eqn:BE-expC2}
\end{equation}
where $A$, $B$, and $C$ are fitting parameters and 
$E_\text{mono}(r_\text{s})$ is the monowire ground state energy per electron at density parameter $r_\text{s}$. The $E_\text{mono}(r_\text{s})$ values are taken from Refs.\ \onlinecite{Lee2011b} and \onlinecite{Ashokan2018a} for low and high density, respectively.  It can be seen from Eq.\ (\ref{eqn:BE-expC2}) that for $d \to 0$,
\begin{equation}
\Delta E \to E_\text{mono}(r_\text{s}/2)-E_\text{mono}(r_\text{s})+\ln(d^2)/(4r_\text{s}),
\end{equation}
and for  $d \to \infty$
\begin{equation}
\Delta E \to \frac{A}{Cd^2}.
\end{equation}
Fitted curves using Eq.\ (\ref{eqn:BE-expC2}) are shown in Fig.\ \ref{fig:BE-ndd} for various values of $r_\text{s}$. The quality of fitting is visible in the curve. The fitted parameters are shown in Table \ref{tab:BE-exp-d}. We have fitted Eq.\ (\ref{eqn:BE-expC2}) to our simulation data using two different methods~\cite{fit}; both yield almost identical results.

\begin{table}[!htbp]
\caption{\label{tab:BE-exp-d}
Values of $A$, $B$ and $C$ in Eq.\ (\ref{eqn:BE-expC2}) are obtained at various $r_\text{s}$ from fitting $\Delta E$ data for values of $d$ from 0.1 to 1 a.u.
}
\begin{ruledtabular}
\begin{tabular}{lccc}
$r_\text{s}$ & $A$ & $B$ & $C$ \\ \hline
0.2  & $-2.1301 \times 10^{17}$ & $-8.4178 \times 10^{17}$ & $1.2827 \times 10^{21}$ \\
0.4  & $-1.3487 \times 10^{17}$ & $6.1654 \times 10^{16}$ & $2.1521 \times 10^{20}$ \\
0.8  & $-4.2687 \times 10^4$ & $3.8129 \times 10^4$ & $1.8522 \times 10^7$ \\
2.0  & $-1.8201 \times 10^2$ & $4.2517 \times 10^2$ & $1.7076 \times 10^4$ \\
10.0 & $-1.7446 \times 10^{-1}$ & $1.3828$ & $1.5523$ \\
20.0 & $-5.7147 \times 10^{-3}$ & $7.1897 \times 10^{-2}$ & $3.9272 \times 10^{-2}$ \\
\end{tabular}
\end{ruledtabular}
\end{table}%

\begin{figure*}[!htbp]
\includegraphics[width=.95\textwidth]{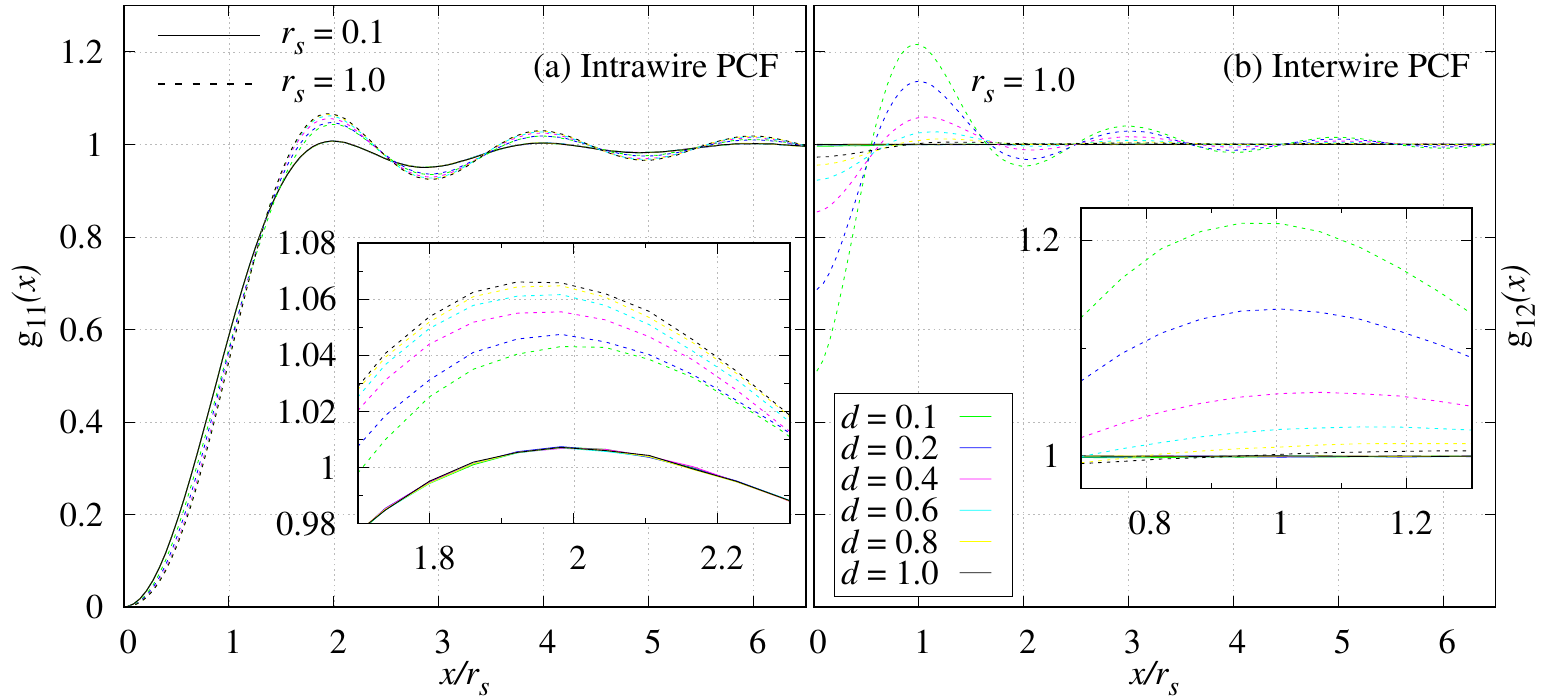}
\caption{\label{fig:g-lrs} Inter- and intra-wire PCFs at different wire separations $d$ for $r_\text{s} = 0.1$ and 1. Solid lines are used for $r_\text{s} = 0.1$ and dashed lines are for $r_\text{s} = 1.0$. The inset shows a magnified view of the first peak of the PCF\@.}
\end{figure*}
\begin{figure}[!htbp]
\centering
\includegraphics[width=.44\textwidth]{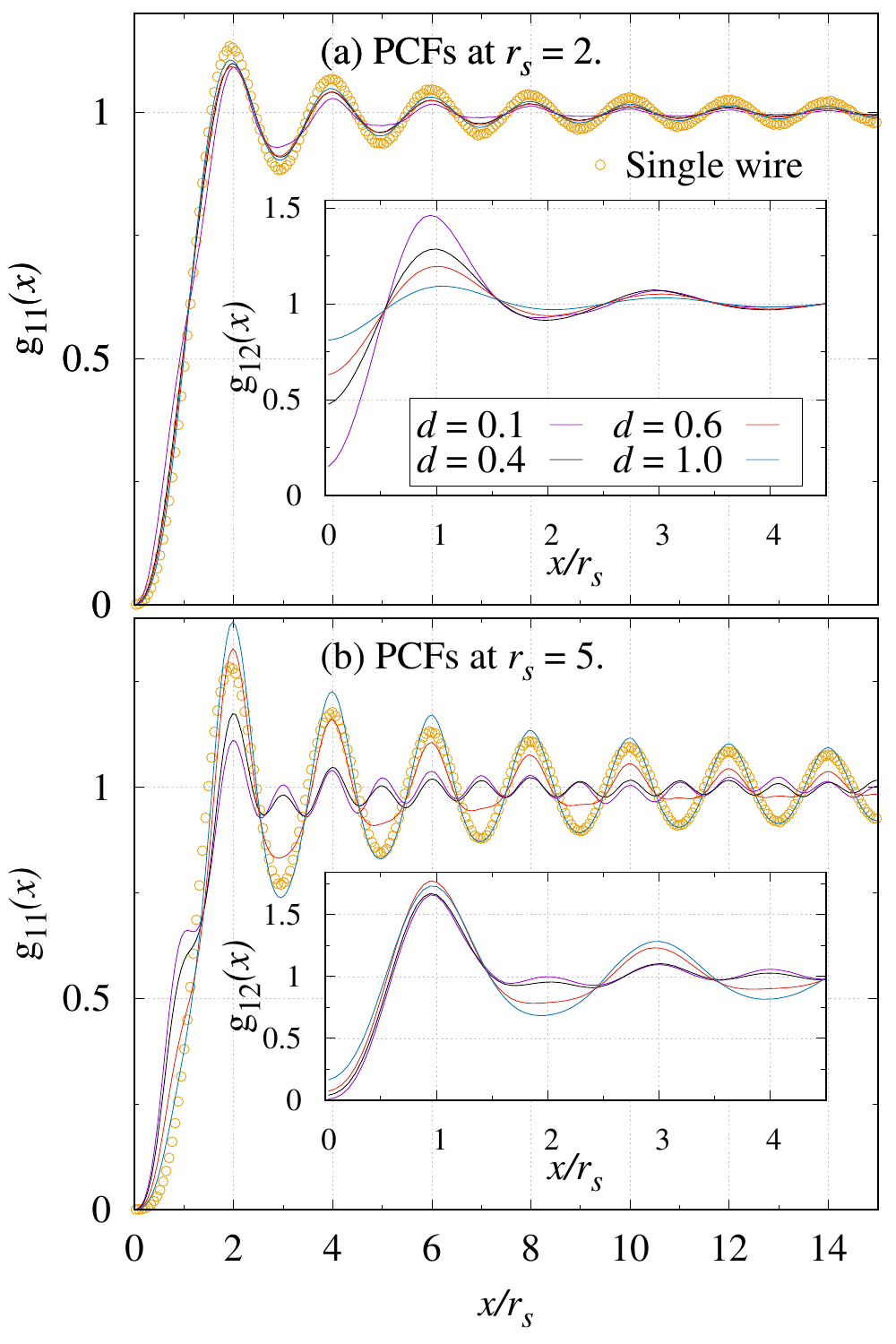}
\caption{\label{fig:g-hrs1} Intra- and inter-wire PCFs at various separations $d$. (a) $g_{11}(r)$ and $g_{12}(r)$ are plotted for the density parameter $r_\text{s} = 2$. (b) $g_{11}(r)$ and $g_{12}(r)$ are plotted for the density parameter $r_\text{s} = 5$. Symbols represent data for a single wire.}
\end{figure}%
\begin{figure}[!htbp]
\centering
\includegraphics[width=.44\textwidth]{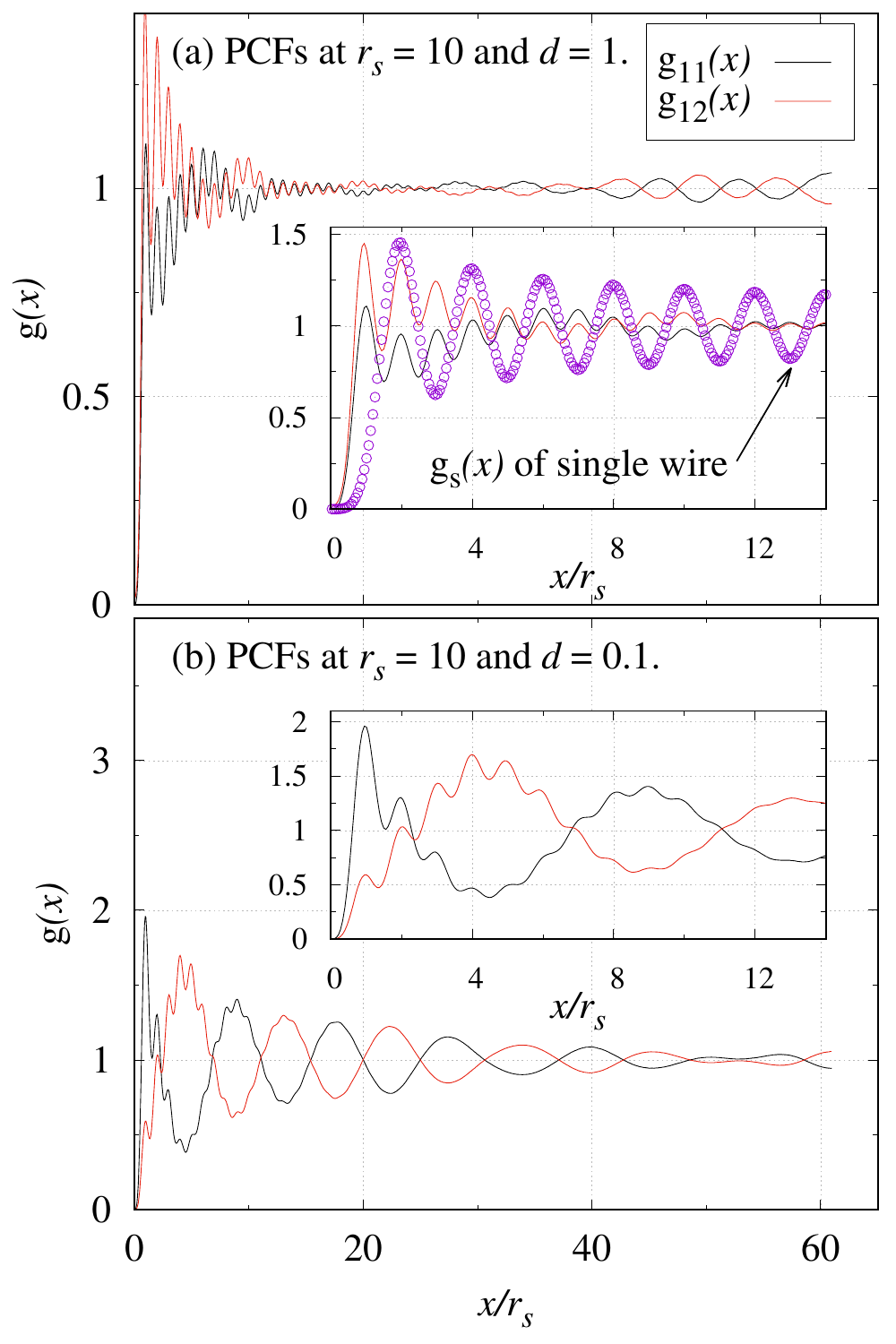}
\caption{\label{fig:g-hrs2} VMC inter- and intra-wire PCFs at $d = 1$ and 0.1 a.u.\ for $r_\text{s} = 10$ are shown in the top and bottom panels, respectively. The inset shows a magnified view. Open circles are used for the single-wire PCF (top panel).}
\end{figure}

\subsection{Pair-correlation functions}\label{sec:IV-B}
The intrawire (parallel-spin) PCF is defined as
\begin{equation}
g_{11}(x)=\dfrac{1}{Ln_{1}^2}\left< \sum_{i\neq j}\delta(x_{i,1}-x_{j,1}-x)\right>,
\end{equation}
where $n_m$ is the average density of electrons in wire $m$ and $L$ is the simulation-cell length. The angular brackets denote an average over the configurations generated by the VMC algorithms. Since both wires are symmetric with respect to the charge and mass of the mobile carriers, $g_{11}$ and $g_{22}$ are equal.  The interwire (antiparallel-spin) PCF may be written as
\begin{equation}
g_{12}(x)=\dfrac{1}{Ln_{1}n_{2}}\left<
\sum_{i,j} \delta(x_{i,1}-x_{j,2}-x)\right>.
\end{equation}

In Figs.\ \ref{fig:g-lrs}(a) and \ref{fig:g-lrs}(b), intra- (same-spin) and inter-wire (opposite-spin) PCFs, respectively, are shown for densities $r_\text{s} = 0.1$ and $r_\text{s} = 1.0$. The intrawire PCF $g_{11}$ shows oscillatory behavior for all the values of interwire separation $d$ that we have considered here. Therefore a significant amount of intrawire electronic correlation is present in the EEBW even at very high densities. Oscillations in $g_{12}$ increase as $d$ is reduced, while oscillations in $g_{11}$ decrease. This reveals that the correlations between electrons in different wires are reinforced and intrawire correlations are suppressed as two wires approach. The first peaks in $g_{11}$ and $g_{12}$ are situated near $r = 2r_\text{s}$ and $r = r_\text{s}$, respectively. Both $g_{11}$ and $g_{12}$ oscillate with a period $2r_\text{s}$. As $d$ is reduced, the first peak of $g_{12}$ rises and shifts towards the origin, while for $g_{11}$ it shrinks and shifts away from origin (see the inset of Fig.\ \ref{fig:g-lrs}), except for $r_\text{s} = 0.1$, where the influence of $d$ is negligibly small. Also note that the value of $g_{12}(r)$ at $r=0$ shifts towards zero as $d$ is reduced, because with decreasing $d$, electrons in different wires repel each other and consequently $g_{12}(0)$ becomes smaller. The value of $g_{12}(0)$ should go to zero as $d\to 0$ at low densities as show in the Fig.\ \ref{fig:g-hrs2}.

The low-density behavior of intra- and inter-wire PCFs is shown in Figs.\ \ref{fig:g-hrs1} and \ref{fig:g-hrs2}. For $r_\text{s}=2$ the behavior of $g_{11}$ and $g_{12}$ presented in Fig.\ \ref{fig:g-hrs1}(a) is similar to that for $r_\text{s}=1$. However, as noticed for $r_\text{s}=5$ in Fig.\ \ref{fig:g-hrs1}(b) a small peak begins to develop in $g_{11}$ at $r=r_\text{s}$ when the interwire distance is reduced to $0.6$ a.u., which keeps rising with further reduction in $d$. At a distance $d = 0.4$ a.u., $g_{11}$ oscillates with a period of $r_\text{s}$ rather than with $r=2r_\text{s}$ as shown in Fig.\ \ref{fig:g-hrs1}(b). Similar to $g_{11}$, also $g_{12}$ begins to oscillate at period $r=r_\text{s}$ for $d \leq 0.4$ a.u., which can be seen in the inset of Fig.\ \ref{fig:g-hrs1}(b). This suggests that when $d$ is large the biwire system is two isolated monowires of number density $N/L$; when $d\to 0$ the biwire system is like a single monowire of number density $2N/L$. The PCFs in Fig.\ \ref{fig:g-hrs2} show strong electronic correlation effects in the low-density regime, where it is seen that at $r_\text{s}=10$ the oscillations in both inter- and intra-wire PCFs are enhanced further. Here, the interwire correlations are comparatively stronger than the intra-wire correlations as the considered range of $d$ is significantly smaller than $r_\text{s}$. From Fig.\ \ref{fig:g-hrs2} it can be seen that PCFs have two kinds of oscillations; the first has a period of $r_\text{s}$ and is enveloped by the second kind of oscillation. This effect arises due to interplay between intra- and interwire correlations.

\begin{figure}[!htbp]
\centering
\includegraphics[width=.48\textwidth]{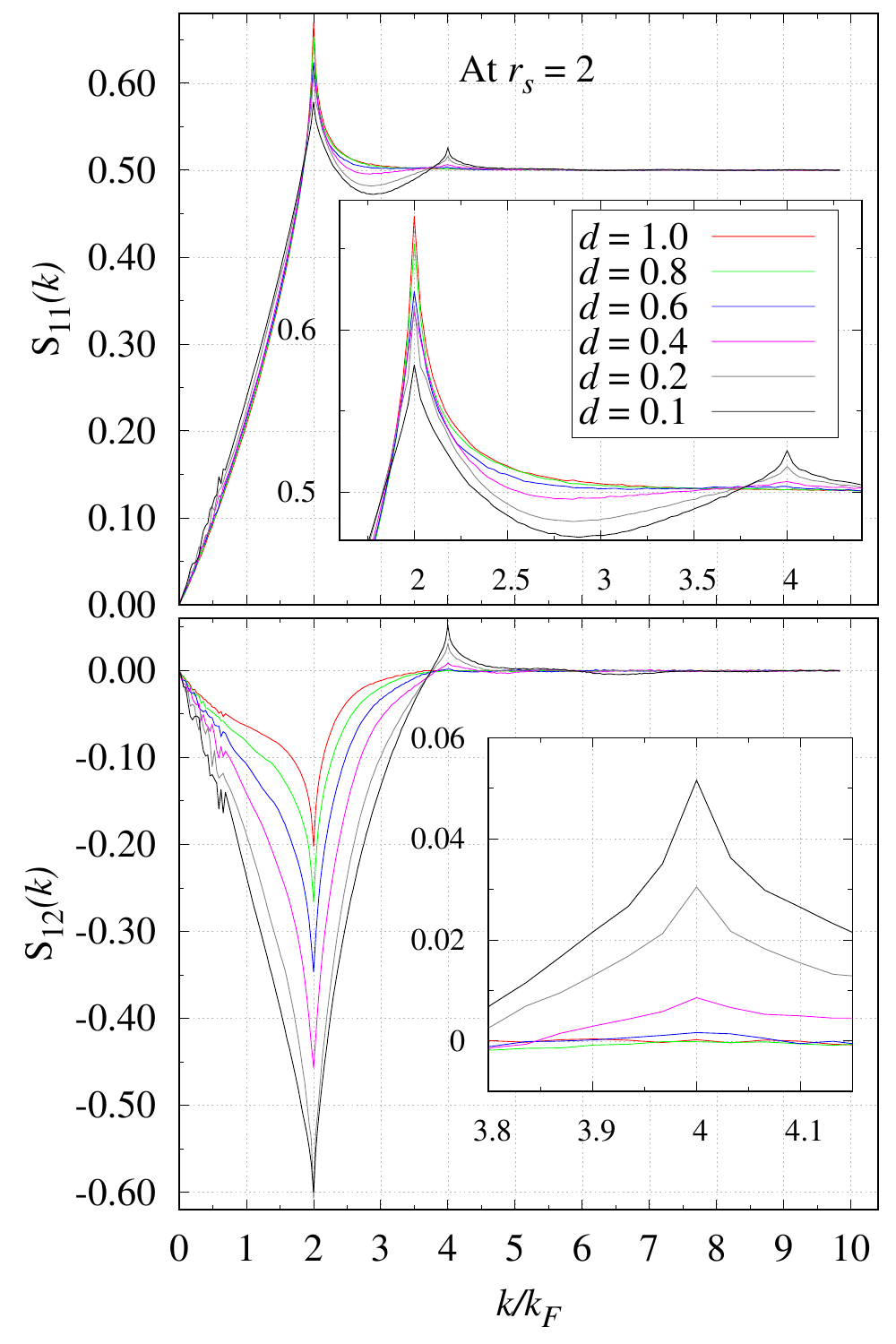}
\caption{\label{fig:S11-S12} Top panel: intrawire SSF $S_{11}$. Bottom panel: interwire SSF $S_{12}$ for various values of $d$ at $r_\text{s}=2$. The inset shows a zoomed-in view of the peaks in the SSF\@.}
\end{figure}%
\begin{figure*}[!htbp]
\includegraphics[width=.98\textwidth]{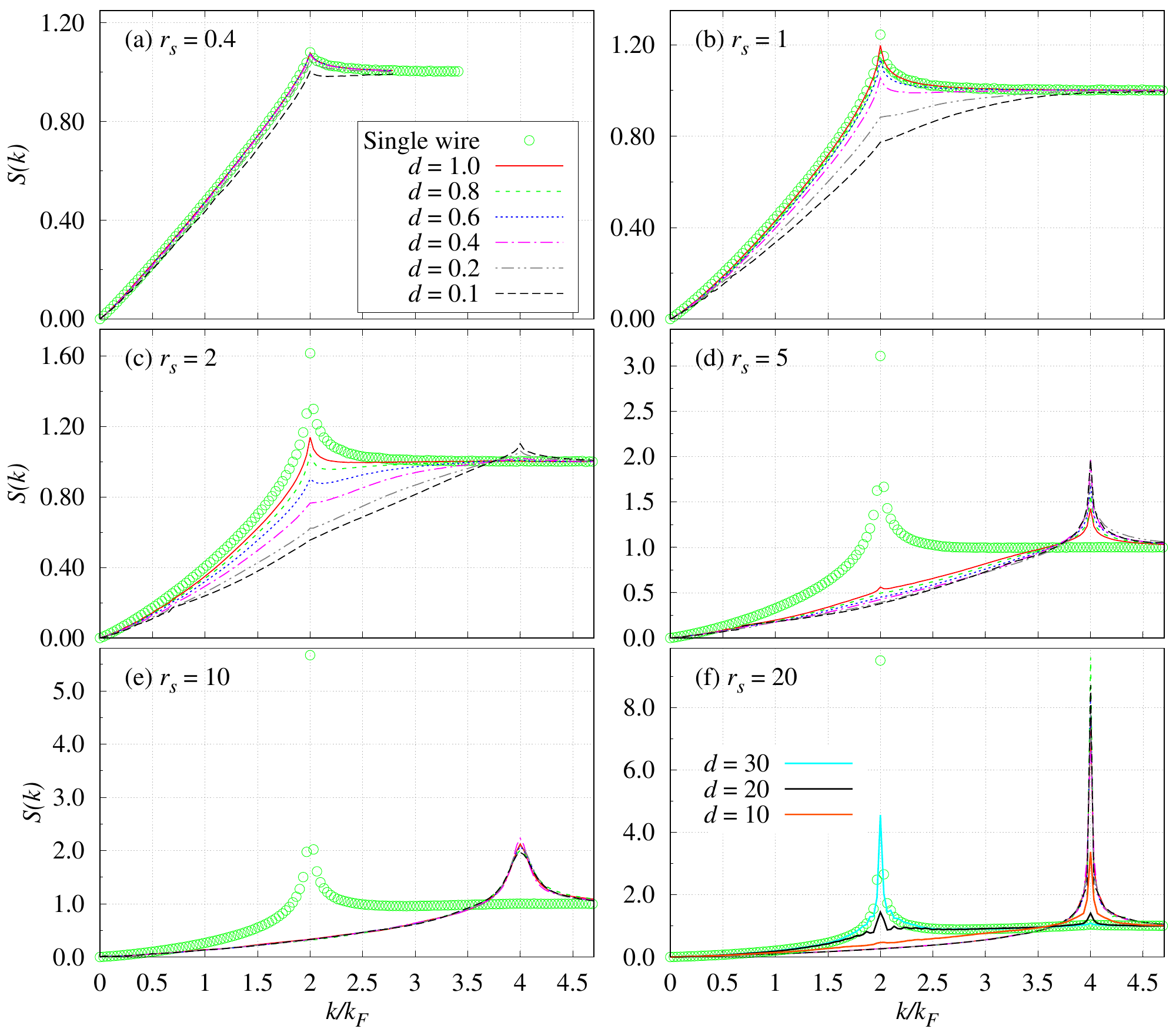}
\caption{\label{fig:SSF} SSFs of EEBWs for various values of $r_\text{s}$ and $d$ are shown with lines. Open circles are used for single-wire SSFs.}
\end{figure*}

\subsection{Static structure factors} 
The SSF is a quantity that can be measured by experiments \cite{Fabbri2015} and contains important information about the structure of the system. For our EEBW system it can be defined as
\begin{equation}
S(k) = 1+\dfrac{2N}{L}\int[g(x)-1]e^{-ikx} \, dx.
\label{eqn:ssf}
\end{equation}
Equation (\ref{eqn:ssf}) involves the density-weighted PCF,
\begin{equation}
   g(x)=\sum_a \sum_b \dfrac{n_a n_b}{n^2} g_{ab},
   \label{eqn:g_sum}
\end{equation}
where $n_a=N_a/L$ is the number density of electrons in wire $a$ and $g_{ab}$ comprises $g_{11}$, $g_{12}$, $g_{21}$, or $g_{22}$. The intrawire $S_{11}(k)$ and interwire $S_{12}(k)$ SSFs are given in Eq.\ (\ref{eqn:ssf}) by using $g_{11}(r)$ and $g_{12}(r)$, respectively. We have obtained $S_{11}(k)$ and $S_{12}(k)$ for all combinations of $r_\text{s}$ and $d$ considered in this paper. $S_{11}(k)$ and $S_{12}(k)$ are shown in the top and bottom panels of Fig.\ \ref{fig:S11-S12}, respectively, for $r_\text{s}=2$ at various values of $d$. The interwire SSF $S_{12}(k)$ is negative in the range of small $k$ values and has a strong peak at $2k_\text{F}$. The $S_{12}(k)$ becomes positive just before $4k_\text{F}$ and a second peak begins to builds up at $4k_\text{F}$ for $d=0.4$ a.u.\ whose height increases as $d$ is reduced further. It is known that the height of the peak in the SSF at $2k_\text{F}$ does not scale as $N$, and hence as $L$, but it appears to be sublinear \cite{Lee2011b, Ashokan2018a}. We have also tested the effect of finite size on the peaks in the SSF, which agrees with previous findings \cite{Lee2011b, Ashokan2018a}. The results are discussed in Sec.\ \ref{sec:IV-FSE} below.

Figure \ref{fig:SSF} shows the SSF calculated by summing over spin pairs, i.e., $S(k)=S_{11}(k)+S_{12}(k)+S_{21}(k)+S_{22}(k)$ using Eq.\ (\ref{eqn:ssf}) and Eq.\ (\ref{eqn:g_sum}) at $r_\text{s}= 0.4$, 1, 2, 5, 10, and 20 for $d \leq 1$ a.u. The SSF of an isolated single wire is also computed for comparison with the SSF of an EEBW, which is shown in Fig.\ \ref{fig:SSF} by open circles. For high densities ($r_\text{s} \leq 1$) the SSF shows a small peak at $2k_\text{F}$ whose height decreases as $d$ becomes smaller, and hence the slope in $S(k)$ decreases for small $k$, as shown in Figs.\ \ref{fig:SSF}(a) and \ref{fig:SSF}(b). Also note that the effect of interwire correlation is more pronounced when $d < r_\text{s}$. The lowering of the height of this peak as two wires approach indicates that the interwire correlation has a strong effect and modifies the overall short-range interactions such that the intrawire correlation is suppressed. Figure \ref{fig:S11-S12} reflects this fact, where one can observe the behavior of the first peak in $S_{11}(k)$ and $S_{12}(k)$ as $d$ changes. For high densities, we can say that $S_{11}(k)$ resembles somewhat the noninteracting structure factor given by the Hartree-Fock approximation. 

As the density is lowered (i.e., $r_\text{s}$ is increased), correlation effects become more important, as depicted in Fig.\ \ref{fig:SSF}(c). There one sees that for $r_\text{s}=2$ a second peak in $S(k)$ begins to appear at $4k_\text{F}$ when $d$ is reduced to $0.2$ a.u., and is enhanced further at $d=0.1$ a.u. No such peak is observed in the single, isolated wire at $r_\text{s}=2$ \cite{Lee2011b}. Lee and Drummond \cite{Lee2011b} found that this peak develops at $4k_\text{F}$ for $r_\text{s} = 15$ a.u.\ in infinitely thin wires using the DMC method. Also notice in Fig.\ \ref{fig:SSF}(c) that the first peak at $2k_\text{F}$ shrinks as $d$ is reduced and completely disappears at $d=0.1$ a.u. For higher values of $r_\text{s}$, the $4k_\text{F}$ peak keeps rising while there is no $2k_\text{F}$ peak for values of $d$ from $1$ to $0.1$ a.u., but it is observed in the single wire and shown by open circles in Fig.\ \ref{fig:SSF}. Figure \ref{fig:SSF}(f) shows that at $r_\text{s}=20$ the peak at $2k_\text{F}$ reappears in the EEBW when $d$ is increased. It is interesting to note that, despite the use of an infinitely thin model, we find a $2k_{\rm F} \rightarrow  4k_{\rm F}$ crossover. This crossover could be due to the presence of the second wire, which provides an extra spin degree of freedom for the strongly-correlated dilute limit $r_\text{s} \gg d$. The peak at $4k_\text{F}$ signals the evolution of the system from two isolated one-component monowires with density parameter $r_\text{s}=2$ to a single two-component monowire with effective density parameter $r_\text{s}=1$, which was also reflected in the PCF (see Sec.\ \ref{sec:IV-B}).

\begin{figure}[!htbp]
\includegraphics[width=.48\textwidth]{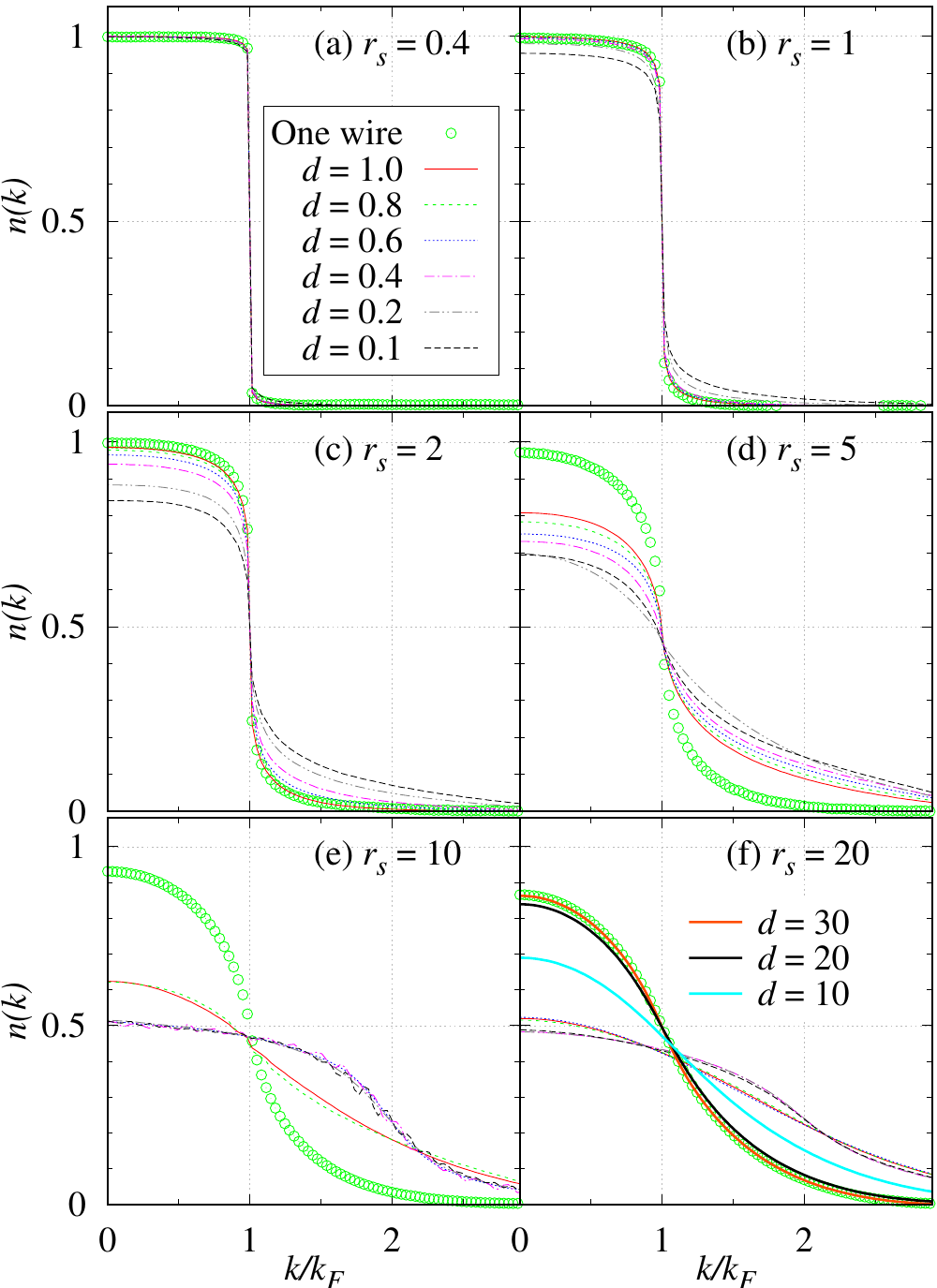}
\caption{\label{fig:md} MD of the EEBW at various values of $r_\text{s}$ and $d$ for $N = 61$. Open circles are for an infinitely thin single wire.}
\end{figure}

\subsection{Momentum densities} 
The MD is calculated from a trial wave function $\Psi_\text{T}$ as
\begin{equation}
n(k)=\dfrac{1}{2\pi}\left<\int\dfrac{\Psi_\text{T}(r)}{\Psi_\text{T}(x_1)}e^{ik(x_1-r)}\,dr \right>,
\label{eqn:md}
\end{equation}
where $\Psi_\text{T}(r)$ is evaluated at $(r, x_2,\ldots,x_N)$. The angular brackets denote the VMC expectation value, obtained as the mean over electron coordinates $(x_1,\ldots,x_N)$ distributed as $|\Psi_\text{T}|^2$. This is an intrawire MD and it will be the same for both wires, although it depends on both the interwire as well as intrawire Coulomb interactions. 

The MD defined through Eq.\ (\ref{eqn:md}) is the Fourier transform of the one-particle density matrix. It is an important quantity from which the TL liquid parameter can be calculated. The MD $n(k)$ gives the occupation of fermionic states with momentum $k$. For a free electron system all the states are completely occupied up to the Fermi energy $E_\text{F}$ at absolute zero temperature, so that $n(k)$ has a discontinuity $Z=1$ at the Fermi momentum $k_\text{F}$. In interacting fermionic systems of dimension higher than one, $n(k)$ still has a discontinuity at the Fermi surface, but its magnitude $Z$ is less than 1. Interacting electrons are now nearly free quasiparticles dressed by density fluctuations \cite{Giamarchi2003}, each of which can move through the Fermi sea by pushing away its neighbors. In contrast, an individual electron in a 1D interacting system cannot move without pushing all the electrons. This results in collective excitations rather than single-particle ones. Thus $n(k)$ has no discontinuity at $k_\text{F}$. TL liquid theory \cite{Luttinger1963, Mattis1965} suggests that $n(k)$ has a power-law behavior close to $k_\text{F}$, which takes the form
\begin{equation}
n(k)=n(k_\text{F})+A[\textrm{sign}(k-k_\text{F})]|k-k_\text{F}|^{\alpha},
\label{eqn:nk_alpha}
\end{equation}
where $n(k_\text{F})$, $A$, and $\alpha$ are constants. We have fitted Eq.\ (\ref{eqn:nk_alpha}) to our MD data to find the value of the exponent $\alpha$.

\begin{figure}[!htbp]
\centering
\includegraphics[width=.48\textwidth]{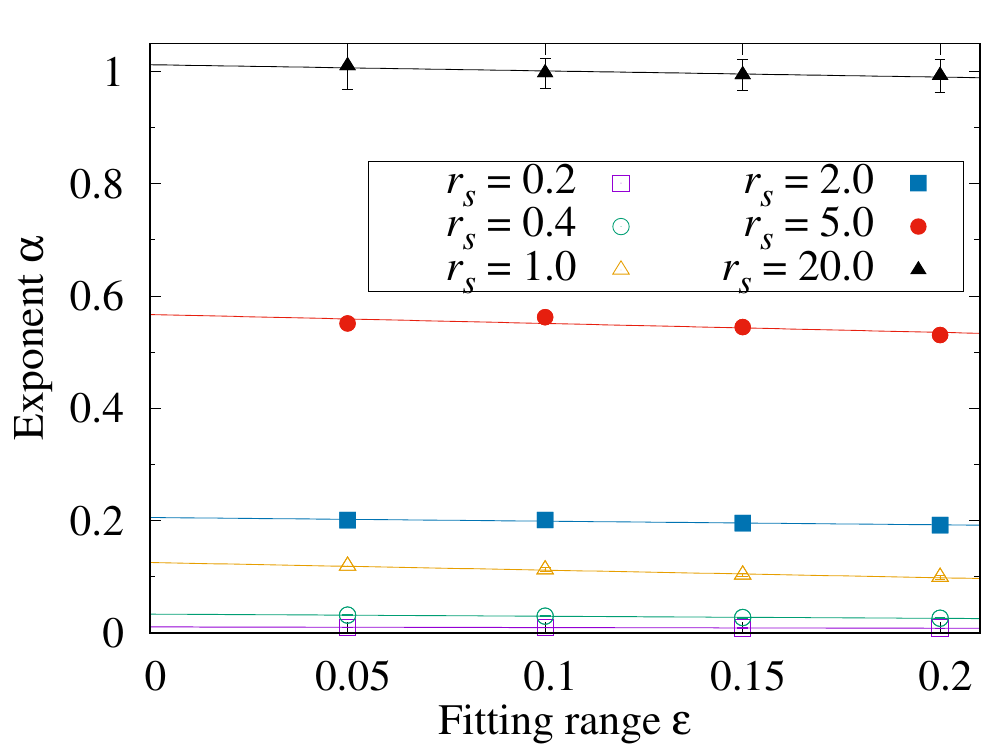}
\caption{\label{fig:alphaE} Exponent $\alpha$ calculated by fitting Eq.\ (\ref{eqn:nk_alpha}) to the MD against the range of data described by $|k - k_\text{F}| < \epsilon k_\text{F}$. The symbols are the fitted exponents and the solid lines are linear fits to the exponents in the region $\epsilon > 0.05$. Here the data are shown for $d=1$ and various values of $r_\text{s}$.}
\end{figure}
\begin{figure}[!htbp]
\centering
\includegraphics[width=.48\textwidth]{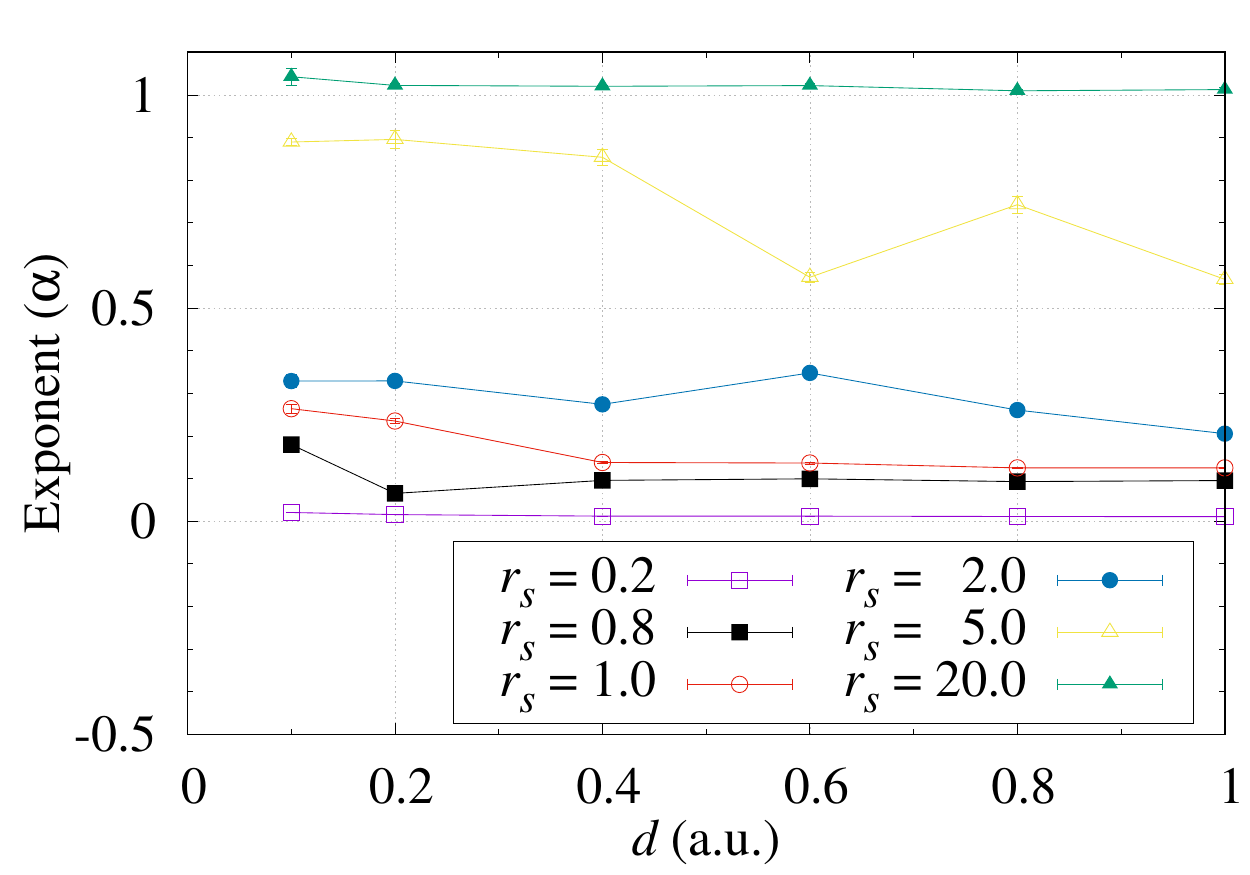}
\caption{\label{fig:alphaD} Extrapolated values of the exponent $\alpha$ vs.\ $d$ at various $r_\text{s}$ as symbols. Solid lines are just to guide the eyes.}
\end{figure}

Figure \ref{fig:md} shows the MD of an EEBW at various values of $r_\text{s}$ and $d$ for $N=61$, including the MD of a single wire (open circles). The effect of interwire correlations is clearly visible for $d \approx r_\text{s}$. As two wires approach from $d=1$ to $0.1$ a.u., the value of $n(k=0)$ reduces from $1$ as seen in Figs.\ \ref{fig:md}(b)--\ref{fig:md}(d). At fixed $d$ the value of $n(k=0)$ also reduces with $r_\text{s}$ as seen in Fig.\ \ref{fig:md}. At very low densities [see Figs.\ \ref{fig:md}(e) and \ref{fig:md}(f)] the value of $n(k=0)$ falls close to $0.5$ for all the values of $d$ we have considered here, as the change in $d$ is very small compared to $r_\text{s}$. However, when $d$ approaches $r_\text{s}$ we can see a change in $n(k)$. At $d=30$ a.u., $n(k)$ for the biwire resembles the single wire. 

The exponent $\alpha$ in Eq.\ (\ref{eqn:nk_alpha}) is found by fitting $n(k)$ within the range $|k - k_\text{F}| < \epsilon k_\text{F}$. The smaller $\epsilon$ is, the narrower the range of $k$ around $k_\text{F}$. Ideally, $\epsilon$ should be zero, as Eq.\ (\ref{eqn:nk_alpha}) is valid for only $k \rightarrow k_\text{F}$.
The value of $\epsilon$ is reduced from 0.2 to 0.05, and at each $\epsilon$ we fit $n(k)$ using Eq.\ (\ref{eqn:nk_alpha}) to find $\alpha(\epsilon)$. These $\alpha(\epsilon)$s are then extrapolated to $\epsilon = 0$ by a linear fit, which is shown in Fig.\ \ref{fig:alphaE} at $d=1$ a.u.\ for various values of $r_\text{s}$.  Figure \ref{fig:alphaE} reveals that in the high-density limit $\alpha$ tends to zero, whereas in the low-density limit $\alpha$ tends to 1. This trend of exponent $\alpha$ is similar to what has been observed for single wires by Lee and Drummond \cite{Lee2011b} and Ashokan \textit{et al.}\ \cite{Ashokan2018a} for low and high densities, respectively. Figure \ref{fig:alphaD} shows the exponent $\alpha$ against the interwire distance $d$ for various values of $r_\text{s}$. It is observed here that $\alpha$ slowly increases as $d$ decreases.

\begin{figure}[!htbp]
\centering
\includegraphics[width=.48\textwidth]{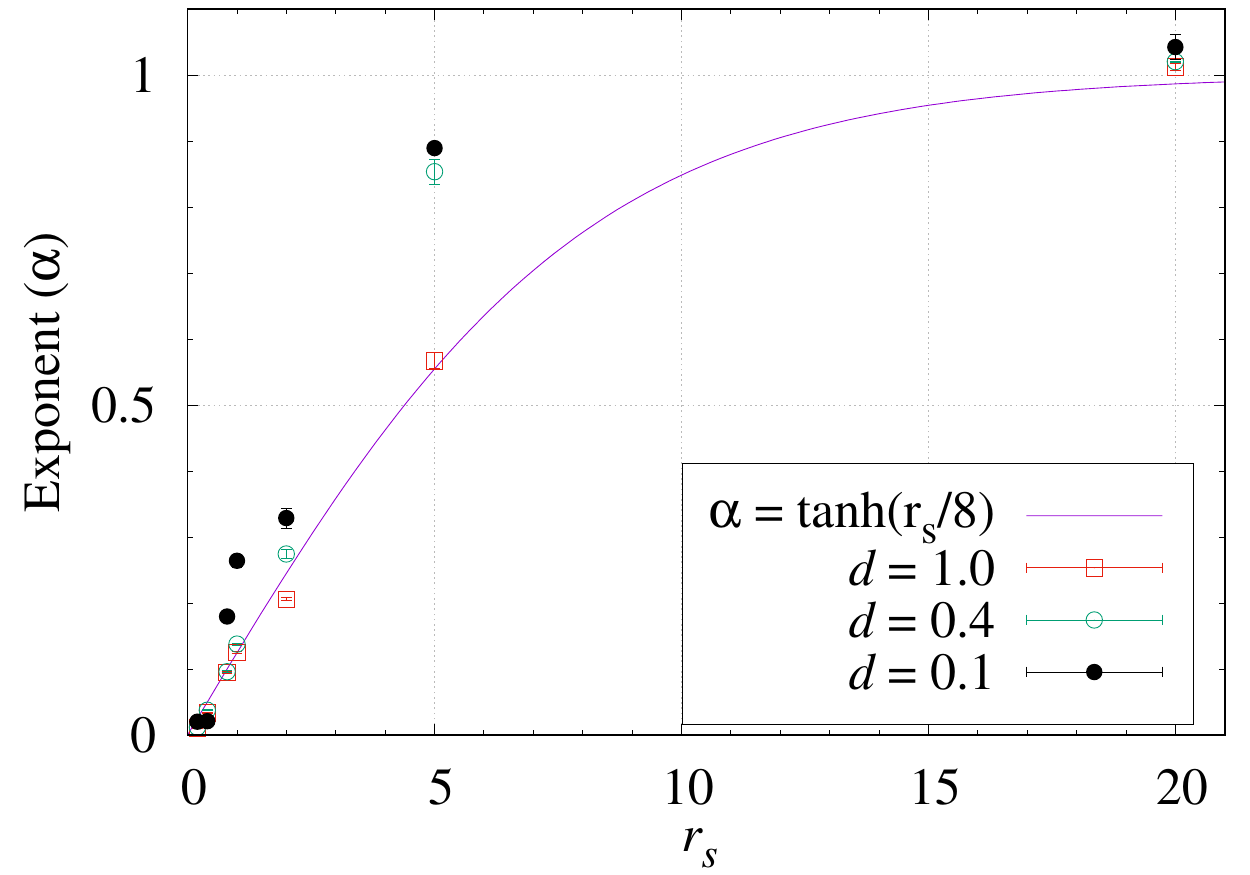}
\caption{\label{fig:alpha_th} VMC exponents $\alpha$ (symbols) plotted against $r_\text{s}$ at various values of $d$. The solid line shows $\alpha$ plotted using Eq.\ (\ref{eqn:alpha_th}) for the infinitely thin single wire, which is close to the VMC EEBW data for $d=1$ a.u.}
\end{figure}

For an isolated, infinitely thin wire the exponent $\alpha$ is reasonably well approximated by the function \cite{Lee2011b}
\begin{equation}
\alpha=\tanh \left(\dfrac{r_\text{s}}{8}\right).
\label{eqn:alpha_th}
\end{equation}
This function is plotted in Fig.\ \ref{fig:alpha_th} vs.\ $r_\text{s}$ with a solid line, to compare with our VMC data (symbols). It is found that $\alpha$ obtained using Eq.\ (\ref{eqn:alpha_th}) passes close to the VMC data for $d=1$ for small $r_\text{s}$. Smaller separations $d$ give larger values of the exponent $\alpha$.

\begin{figure}[!htbp]
\centering
\includegraphics[width=.48\textwidth]{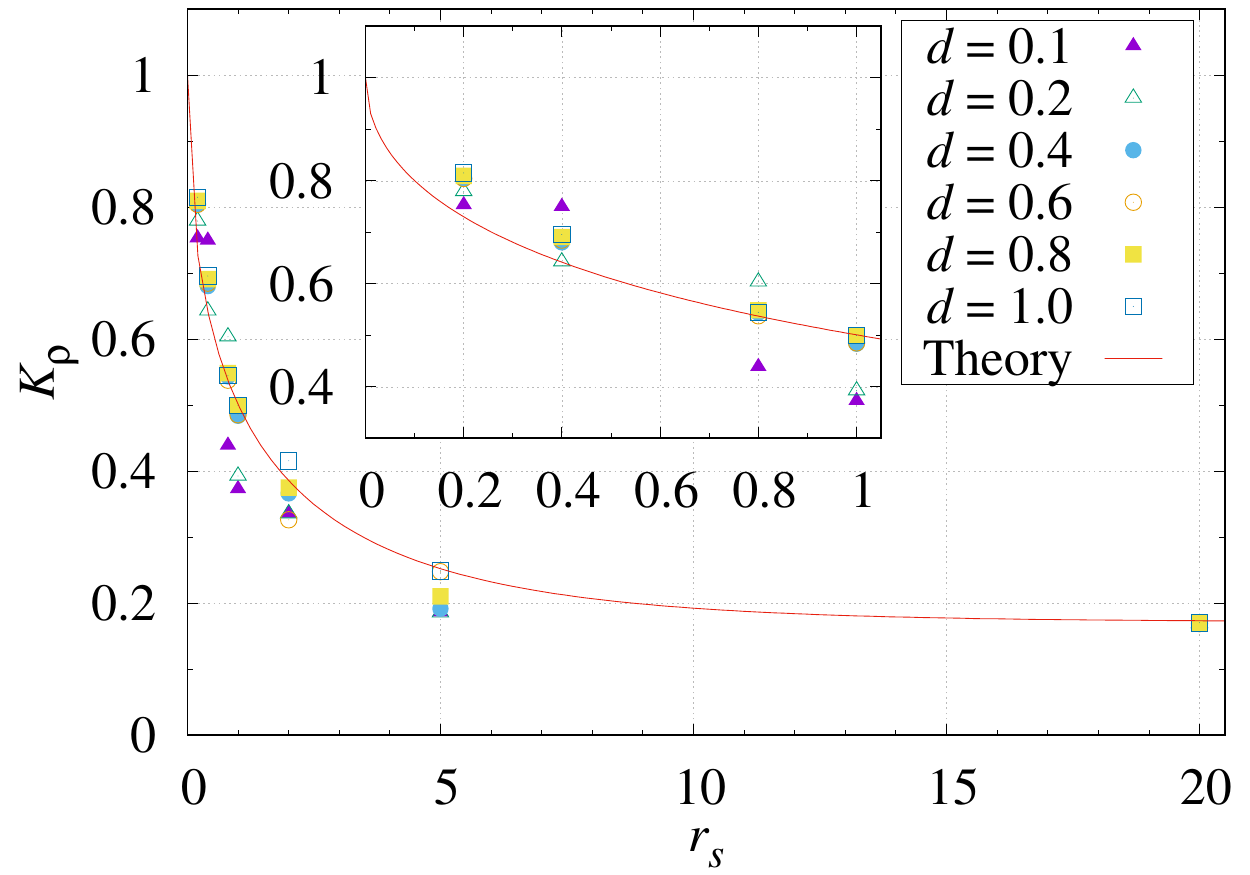}
\caption{\label{fig:K_rho} TL interaction parameter $K_{\rho}$ plotted against $r_\text{s}$ for various wire separations $d$ for EEBWs computed using Eq.\ (\ref{eqn:TLk2}). The solid line shows the theoretical result for an isolated, infinitely thin single wire obtained using Eq.\ (\ref{eqn:TLk3}), which passes close to larger-$d$ EEBW data.}
\end{figure}

Within the TL liquid theory the exponent $\alpha$ is related to the interaction parameter \cite{Schulz1990} $K_{\rho}$ by
\begin{equation}
\alpha=\dfrac{1}{4}\left(K_{\rho}+\dfrac{1}{K_{\rho}}-2 \right).
\label{eqn:TLk}
\end{equation}
By rearranging the above Eq.\ (\ref{eqn:TLk}), the Luttinger parameter $K_{\rho}$ can be written in terms of $\alpha$ as
\begin{equation}
K_{\rho}=1+2\alpha-2\sqrt{\alpha+\alpha^2}
\label{eqn:TLk2}
\end{equation}
Note that $K_{\rho} = 1$ for noninteracting particles, $K_{\rho} > 1$ for attractive interactions, and $0<K_{\rho}<1$ for repulsive interactions. For strong repulsive interactions $K_{\rho} \ll 1$. Therefore, $K_{\rho}$ gives a quantitative value of the correlation strength. We calculated $K_{\rho}$ in Eq.\ (\ref{eqn:TLk2}) by using values of the extrapolated exponent $\alpha$ obtained at various values of $r_\text{s}$ and $d$. The results are plotted in Fig.\ \ref{fig:K_rho} against $r_\text{s}$ for various values of $d$ indicated by symbols. The inset shows the same data for small $r_\text{s}$. Further, $K_{\rho}$ can be written in terms of $r_\text{s}$ by using Eq.\ (\ref{eqn:alpha_th}) in Eq.\ (\ref{eqn:TLk2}) as
\begin{equation}
K_{\rho}=1+2\tanh \left(\dfrac{r_\text{s}}{8}\right)-2\sqrt{\tanh \left(\dfrac{r_\text{s}}{8}\right)+\tanh^2\left(\dfrac{r_\text{s}}{8}\right)}.
\label{eqn:TLk3}
\end{equation}
Using Eq.\ (\ref{eqn:TLk3}), $K_{\rho}$ is plotted by a solid line in Fig.\ \ref{fig:K_rho}. Note that Eq.\ (\ref{eqn:TLk3}) is valid for an isolated single wire; similar $K_\rho$ values are obtained for the $d=1$ a.u.\ EEBW\@.

\section{Conclusions}\label{sec:V}
In this paper we report the ground-state properties of an infinitely thin quantum EEBW system for various electron densities ($r_\text{s}$) and interwire separations ($d$). We use the VMC method to calculate the ground-state energy, PCF, SSF, and MD at three different system sizes. VMC ground-state energies are extrapolated to the thermodynamic limit. The $4k_{\rm F}$ peak of the SSF has a significant finite-size scaling, although sublinear.
For the other observables we find a negligible finite-size effect; hence they are presented as obtained at the largest system size studied. Using the extrapolated ground-state energy, we have computed the correlation energy and the interaction energy per electron for the EEBW system in the thermodynamic limit. We find that the interaction energy increases logarithmically for $d \ll r_\text{s}$ and decreases as a power law with an exponent of $-2$ for $d \gg r_\text{s}$. The correlation energy follows the same trend with $d$ as the interaction energy because the correlation energy of a biwire is the sum of the correlation energy of a single wire and the interaction energy of the biwire. Both inter- and intra-wire PCFs show oscillatory behavior at all densities considered here. As two wires approach each other at a given density parameter $r_\text{s}$, the oscillations in the interwire PCF are enhanced while oscillations in the intrawire PCF are suppressed for $d<r_\text{s}$. This suggests that the interwire correlation increases and intrawire correlation decreases as the wire separation is decreased.  At high densities $r_\text{s} \leq 2$, both PCFs oscillate with a period of $2r_\text{s}$ at all wire separations $d$ considered in this study. However, when $d$ is reduced to $0.4$ a.u.\ at $r_\text{s}=5$, both PCFs begin to oscillate with a period of $r_\text{s}$ instead of $2r_\text{s}$. Their amplitudes increase as $d$ is reduced further. 
This indicates that the system evolves into a single monowire with double the electron density from two isolated monowires as $d$ is reduced from infinity to $0$. This result is also confirmed by our SSF data, which shows a sharp peak at $4k_\text{F}$ that corresponds to a distance $r_\text{s}$ in real space [i.e.\ $r=2\pi/(4k_\text{F})=r_\text{s}$, where $k_\text{F}=\pi/2r_\text{s}$]. At lower $r_\text{s}$ the SSF shows a peak at $2k_\text{F}$ only. The height of this peak decreases as $d$ is reduced. A second peak starts to appear at $4k_\text{F}$ when $d = 0.2$ a.u.\ and $r_\text{s} = 2$. For higher $r_\text{s}$, the first peak completely disappears and the height of the second peak keeps increasing with $d$ and $r_\text{s}$.

The MD $n(k)$ shows TL liquid behavior, as $n(k)$ follows a power law in $|k-k_{\rm F}|$ near $k_\text{F}$. The value of $n(k=0)$ reduces and reaches $0.5$ as $d$ decreases and as the density decreases, which is compensated by an increase in $n(k)$ beyond $k_\text{F}$. We have obtained the TL liquid exponent $\alpha$ by fitting the MD data near $k_\text{F}$. The values of the exponent $\alpha$ shift towards $1$ as the density is lowered and towards $0$ if the density is increased. At fixed $r_\text{s}$, the exponent $\alpha$ increases slowly as $d$ is decreased. Using the exponent $\alpha$ we have calculated the TL liquid interaction parameter $K_{\rho}$.
We find that at a fixed density, the value of $K_{\rho}$ reduces as the interwire distance decreases. At fixed $d$, the value of $K_{\rho}$ reduces as the electron density decreases. As one of the most important conclusions from the EEBW system, we consider that the MD data clearly indicate TL liquid behavior, in spite of the extra interwire interaction between the electrons.

\begin{acknowledgments}
The authors (R.O.S.\ and K.N.P.)\ acknowledge financial support from The National Academy of Sciences, India (NASI)\@. V.A.\ acknowledges support in the form of DST-SERB Grant No.\ EEQ/2019/000528. Computing resources were provided by the WWU IT of M\"{u}nster University (PALMA-II HCP cluster) and Campus Cluster of M\"{u}nster University of Applied Sciences. R.O.S.\ and K.N.P.\ thank Markus Christian Gilbert and Holger Angenent for their support regarding HPC clusters which made this work possible during period of the COVID-19 pandemic.
\end{acknowledgments}


\appendix   

\section{Table of energy data}\label{ap-i}

Table \ref{tab:table1} shows the VMC energies calculated for different system sizes and various values of $r_\text{s}$ and $d$. $E_{\infty}$ gives the ground-state energy, extrapolated to the thermodynamic limit, obtained by fitting Eq.\ (\ref{eqn:Ethrmdnm}). $\Delta E$ and $E_\text{c}$ are the interaction energy per electron and the correlation energy per electron obtained from $E_{\infty}$.


\squeezetable
\begingroup
\begin{table*}[!htbp]
\caption{\label{tab:table1}VMC ground-state energy in a.u.\ per electron [$E(N)$] for $N = 21$, 41, and 61 at various value of $r_\text{s}$ and $d$ for an EEBW system. The $E_{\infty}$ gives the ground-state energy per electron extrapolated to thermodynamic limit. $\Delta E$ and $E_\text{c}$ are the interaction energy and the correlation energy per electron, respectively.}
\begin{ruledtabular}
\begin{tabular}{lcccccc}
($r_\text{s}, d$) & $E(21)$ & $E(41)$& $E(61)$& $E_{\infty}$ & $\Delta E$& $E_\text{c}$\\ \hline
(0.1, 0.1)& 50.146954(5) & 50.222000(4) & 50.236785(6) & 50.24879(9) & $-$0.00477(9) & $-$0.03147(9) \\
(0.1, 0.2)& 50.150400(1) & 50.225487(1) & 50.240269(1) & 50.25229(8) & $-$0.00127(8) & $-$0.02798(8) \\
(0.1, 0.4)& 50.151177(1) & 50.226309(1) & 50.2410973(9) & 50.25312(8) & $-$0.00044(8) & $-$0.02714(8) \\
(0.1, 0.6)& 50.151300(1) & 50.226452(1) & 50.2412409(9) & 50.25327(8) & $-$0.00029(8) & $-$0.02700(8) \\
(0.1, 0.8)& 50.151331(2) & 50.226495(1) & 50.241293(1) & 50.25332(8) & $-$0.00024(8) & $-$0.02694(8) \\
(0.1, 1.0)& 50.151344(1) & 50.226515(1) & 50.241313(1) & 50.25335(8) & $-$0.00021(8) & $-$0.02692(8) \\

(0.2, 0.1)& 13.055245(8) & 13.075366(7) & 13.07960(1) & 13.0827(2) & $-$0.0178(2) & $-$0.0437(2) \\
(0.2, 0.2)& 13.068909(3) & 13.089063(1) & 13.093080(3) & 13.09629(5) & $-$0.00422(5) & $-$0.03014(5) \\
(0.2, 0.4)& 13.0720588(6) & 13.0922567(4) & 13.0962576(8) & 13.09948(4) & $-$0.00103(4) & $-$0.02695(4) \\
(0.2, 0.6)& 13.0725809(6) & 13.0928003(4) & 13.096814(2) & 13.10004(4) & $-$0.00047(4) & $-$0.02639(4) \\
(0.2, 0.8)& 13.0727477(6) & 13.0929807(4) & 13.0969900(8) & 13.10022(4) & $-$0.00029(4) & $-$0.02621(4) \\
(0.2, 1.0)& 13.0728195(6) & 13.0930607(4) & 13.0970708(8) & 13.10030(4) & $-$0.00021(4) & $-$0.02613(4) \\

(0.4, 0.1)& 3.03311(1) & 3.03911(1) & 3.04044(1) & 3.04136(9) & $-$0.06066(9) & $-$0.08521(9) \\
(0.4, 0.2)& 3.078375(5) & 3.084075(5) & 3.085226(5) & 3.08613(2) & $-$0.01588(2) & $-$0.04044(2) \\
(0.4, 0.4)& 3.090797(1) & 3.0964498(8) & 3.097585(1) & 3.09848(2) & $-$0.00353(2) & $-$0.02809(2) \\
(0.4, 0.6)& 3.0928412(4) & 3.0985146(4) & 3.0996452(3) & 3.10055(1) & $-$0.00146(1) & $-$0.02602(1) \\
(0.4, 0.8)& 3.0934943(3) & 3.0991783(2) & 3.1003123(2) & 3.10122(1) & $-$0.00079(2) & $-$0.02535(1) \\
(0.4, 1.0)& 3.0937761(3) & 3.0994693(3) & 3.1006048(2) & 3.10151(1) & $-$0.00050(2) & $-$0.02506(1) \\

(0.8, 0.1)& 0.30509(1) & 0.30651(1) & 0.307126(7) & 0.3072(2) & $-$0.1579(2) & $-$0.1803(2) \\
(0.8, 0.2)& 0.407929(7) & 0.409603(9) & 0.409936(4) & 0.410203(4) & $-$0.054952(4) & $-$0.077313(4) \\
(0.8, 0.4)& 0.449477(2) & 0.451173(4) & 0.451454(1) & 0.45174(3) & $-$0.01341(3) & $-$0.03577(3) \\
(0.8, 0.6)& 0.457647(1) & 0.459312(2) & 0.4596454(6) & 0.459910(5) & $-$0.005245(5) & $-$0.027606(5) \\
(0.8, 0.8)& 0.4602188(6) & 0.4618881(6) & 0.4622234(3) & 0.462488(6) & $-$0.002667(6) & $-$0.025028(6) \\
(0.8, 1.0)& 0.4612897(3) & 0.4629669(4) & 0.4633028(2) & 0.463569(5) & $-$0.001586(5) & $-$0.023947(5) \\

(1.0, 0.1)& $-$0.04392(2) & $-$0.043930(7) & $-$0.043771(6) & $-$0.04382(9) & $-$0.19801(9) & $-$0.21945(9) \\
(1.0, 0.2)& 0.075728(7) & 0.077025(8) & 0.077256(4) & 0.07747(1) & $-$0.07672(1) & $-$0.09816(1) \\
(1.0, 0.4)& 0.132008(3) & 0.133128(3) & 0.133363(2) & 0.133538(9) & $-$0.020651(9) & $-$0.042095(9) \\
(1.0, 0.6)& 0.144508(1) & 0.145626(2) & 0.1458558(9) & 0.146032(7) & $-$0.008157(7) & $-$0.029601(7) \\
(1.0, 0.8)& 0.1485781(8) & 0.1497088(8) & 0.1499387(4) & 0.150117(5) & $-$0.004071(6) & $-$0.025515(5) \\
(1.0, 1.0)& 0.1502710(5) & 0.1514101(5) & 0.1516395(2) & 0.151820(4) & $-$0.002368(4) & $-$0.023812(4) \\

(2.0, 0.1)& $-$0.499396(5) & $-$0.498767(4) & $-$0.498530(4) & $-$0.49846(7) & $-$0.29226(7) & $-$0.31019(7) \\
(2.0, 0.2)& $-$0.367626(4) & $-$0.367240(4) & $-$0.367101(3) & $-$0.36706(4) & $-$0.16086(4) & $-$0.17878(4) \\
(2.0, 0.4)& $-$0.271756(3) & $-$0.271209(4) & $-$0.271211(2) & $-$0.27109(6) & $-$0.06489(6) & $-$0.08281(6) \\
(2.0, 0.6)& $-$0.237874(2) & $-$0.237468(3) & $-$0.237412(1) & $-$0.23734(1) & $-$0.03114(1) & $-$0.04906(1) \\
(2.0, 0.8)& $-$0.223321(1) & $-$0.222964(2) & $-$0.2228821(9) & $-$0.222829(7) & $-$0.016628(7) & $-$0.034550(7) \\
(2.0, 1.0)& $-$0.216339(1) & $-$0.215983(1) & $-$0.2159077(6) & $-$0.215852(3) & $-$0.009651(3) & $-$0.027573(3) \\

(5.0, 0.1)& $-$0.460006(1) & $-$0.459622(1) & $-$0.459032(2) & $-$0.4591(3) & $-$0.2552(3) & $-$0.2675(3) \\
(5.0, 0.2)& $-$0.392209(1) & $-$0.391788(1) & $-$0.3913920(9) & $-$0.3914(2) & $-$0.1875(2) & $-$0.1998(2) \\
(5.0, 0.4)& $-$0.326161(1) & $-$0.3259513(9) & $-$0.3257966(8) & $-$0.32580(7) & $-$0.12187(7) & $-$0.13418(7) \\
(5.0, 0.6)& $-$0.289991(1) & $-$0.2898839(9) & $-$0.2898403(8) & $-$0.28983(1) & $-$0.08590(1) & $-$0.09822(1) \\
(5.0, 0.8)& $-$0.266561(1) & $-$0.266459(2) & $-$0.2663790(8) & $-$0.26638(3) & $-$0.06245(3) & $-$0.07477(3) \\
(5.0, 1.0)& $-$0.250357(1) & $-$0.250224(1) & $-$0.2501940(8) & $-$0.250174(3) & $-$0.046242(3) & $-$0.058560(3) \\

(10.0, 0.1)& $-$0.3189958(9) & $-$0.318522(7) & $-$0.3184203(6) & $-$0.318347(5) & $-$0.175478(5) & $-$0.183770(5) \\
(10.0, 0.2)& $-$0.2844301(8) & $-$0.2842379(5) & $-$0.2840481(5) & $-$0.28407(9) & $-$0.14120(9) & $-$0.14949(9) \\
(10.0, 0.4)& $-$0.2500071(4) & $-$0.2498683(4) & $-$0.2494844(5) & $-$0.2496(2) & $-$0.1067(2) & $-$0.1150(2) \\
(10.0, 0.6)& $-$0.2300087(3) & $-$0.2298816(3) & $-$0.2292813(4) & $-$0.2294(3) & $-$0.0866(3) & $-$0.0949(3) \\
(10.0, 0.8)& $-$0.2161269(3) & $-$0.2158759(3) & $-$0.2152083(6) & $-$0.2154(4) & $-$0.0725(4) & $-$0.0808(4) \\
(10.0, 1.0)& $-$0.2054852(3) & $-$0.2052825(3) & $-$0.2043781(5) & $-$0.2046(5) & $-$0.0617(5) & $-$0.0700(5) \\

(15.0, 0.1)& $-$0.2450628(7) & $-$0.2450197(4) & $-$0.2450174(2) & $-$0.245009(3) & $-$0.134542(3) & $-$0.140861(3) \\
(15.0, 0.2)& $-$0.2219246(3) & $-$0.2219153(2) & $-$0.2219253(1) & $-$0.221920(7) & $-$0.111453(7) & $-$0.117773(7) \\
(15.0, 0.4)& $-$0.1988801(2) & $-$0.1988425(2) & $-$0.19883510(9) & $-$0.19882906(1) & $-$0.08836229(2) & $-$0.09468170(1) \\
(15.0, 0.6)& $-$0.1854460(2) & $-$0.1853468(2) & $-$0.18533123(9) & $-$0.185314(2) & $-$0.074847(2) & $-$0.081167(2) \\
(15.0, 0.8)& $-$0.1759240(2) & $-$0.1757860(2) & $-$0.17575917(9) & $-$0.17573698(3) & $-$0.06527022(3) & $-$0.07158963(3) \\
(15.0, 1.0)& $-$0.1685954(2) & $-$0.1683892(2) & $-$0.1683339(1) & $-$0.168305(9) & $-$0.057839(9) & $-$0.064158(9) \\

(20.0, 0.1)& $-$0.2004362(6) & $-$0.2004050(2) & $-$0.2003922(2) & $-$0.200389(4) & $-$0.109612(4) & $-$0.114744(4) \\
(20.0, 0.2)& $-$0.1831133(1) & $-$0.1830750(2) & $-$0.1830678(1) & $-$0.1830616(2) & $-$0.0922838(2) & $-$0.0974163(2) \\
(20.0, 0.4)& $-$0.1658005(1) & $-$0.1657685(1) & $-$0.16576161(8) & $-$0.1657567(4) & $-$0.0749789(4) & $-$0.0801114(4) \\
(20.0, 0.6)& $-$0.1556847(2) & $-$0.15564145(7) & $-$0.1556384(1) & $-$0.155630(3) & $-$0.064852(3) & $-$0.069984(3) \\
(20.0, 0.8)& $-$0.1485275(1) & $-$0.1484434(1) & $-$0.1484347(1) & $-$0.148419(4) & $-$0.057641(4) & $-$0.062773(4) \\
(20.0, 1.0)& $-$0.1429959(1) & $-$0.1429221(1) & $-$0.14289739(8) & $-$0.142889(6) & $-$0.052111(6) & $-$0.057244(6)
\end{tabular}
\end{ruledtabular}
\end{table*}%
\endgroup

\section{Finite-size effects\label{sec:IV-FSE}}

\begin{figure}[!htbp]
\centering
\includegraphics[width=.45\textwidth]{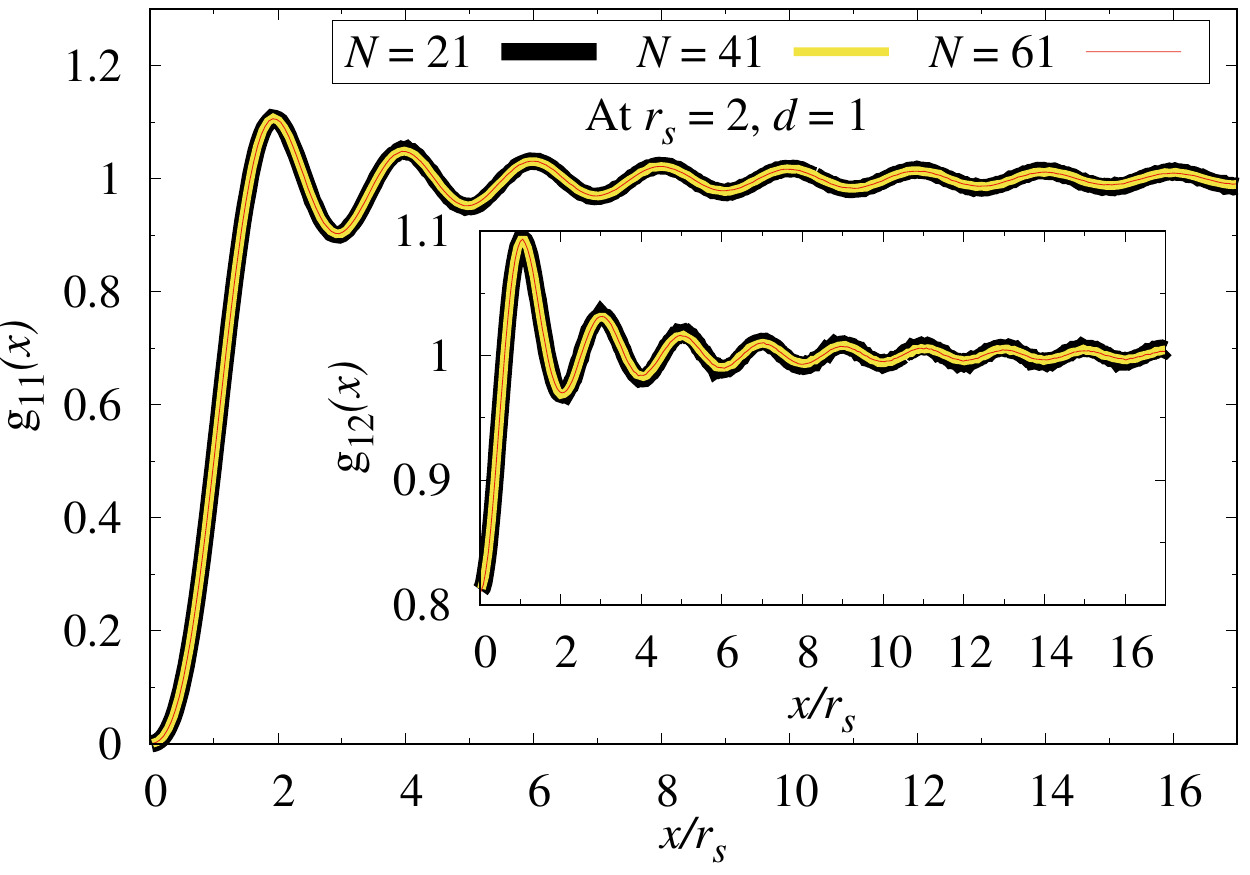}
\caption{\label{fig:g-N} Intra- and inter-wire PCFs as functions of system size $N$ for $r_\text{s} = 2$ a.u.\ and $d = 1$ a.u. The main plot shows the intrawire pair correlation function and the inset shows the interwire pair correlation function. The plot for $N=41$ overlaps on $N=21$ and the one for $N=61$ overlaps on $N=41$. }
\end{figure}

In this section we investigate the effects of finite system sizes on the PCF, SSF, and MD\@. Figure \ref{fig:g-N} shows the intrawire PCF as a function of system size at $d=1$ a.u.\ and $r_\text{s}=2$. We find that the finite-size effect is negligibly small, because it is observed that the PCFs for $N=21$, 41, and 61 overlap.

\begin{figure}[!htbp]
\centering
\includegraphics[width=.45\textwidth]{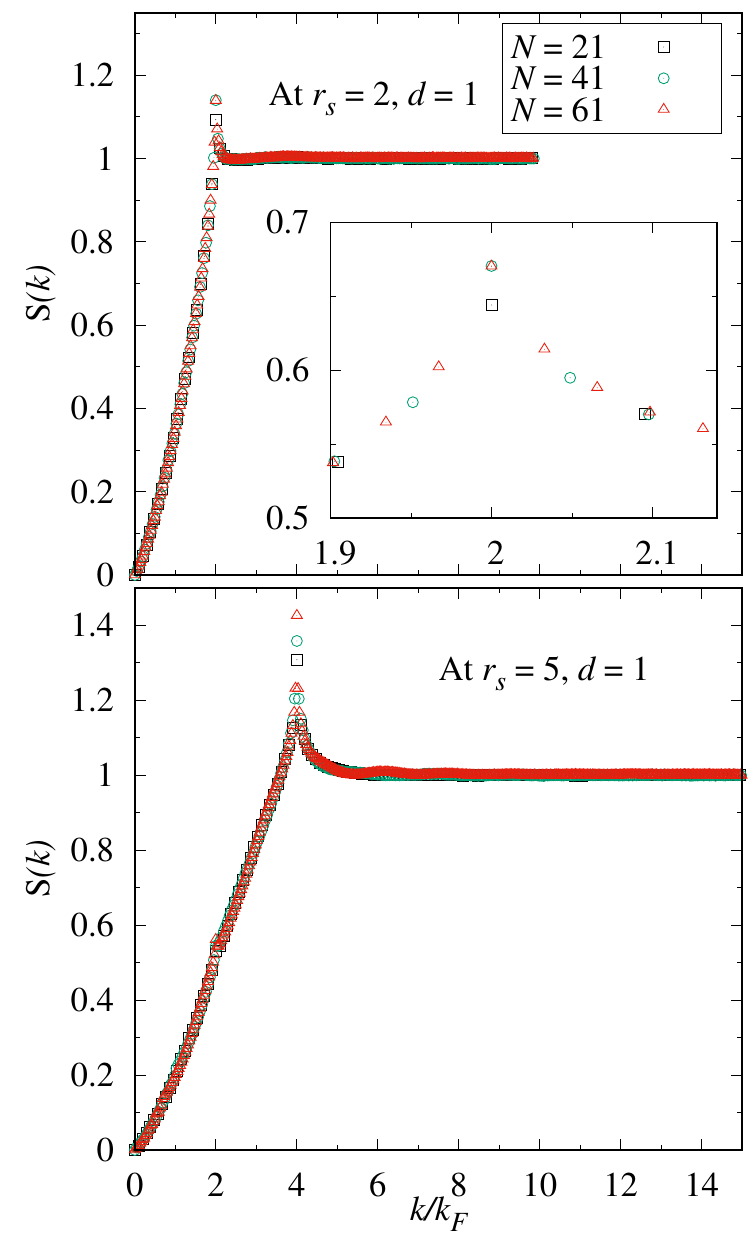}
\caption{\label{fig:SSF-N} SSFs at different system sizes at $d = 1$ a.u.\ for $r_\text{s} = 2$ (top panel) and 5 (bottom panel). The inset in the top panel shows a zoomed-in view of the same data near $2k_\text{F}$.}
\end{figure}

\begin{figure}[!htbp]
\centering
\includegraphics[width=.48\textwidth]{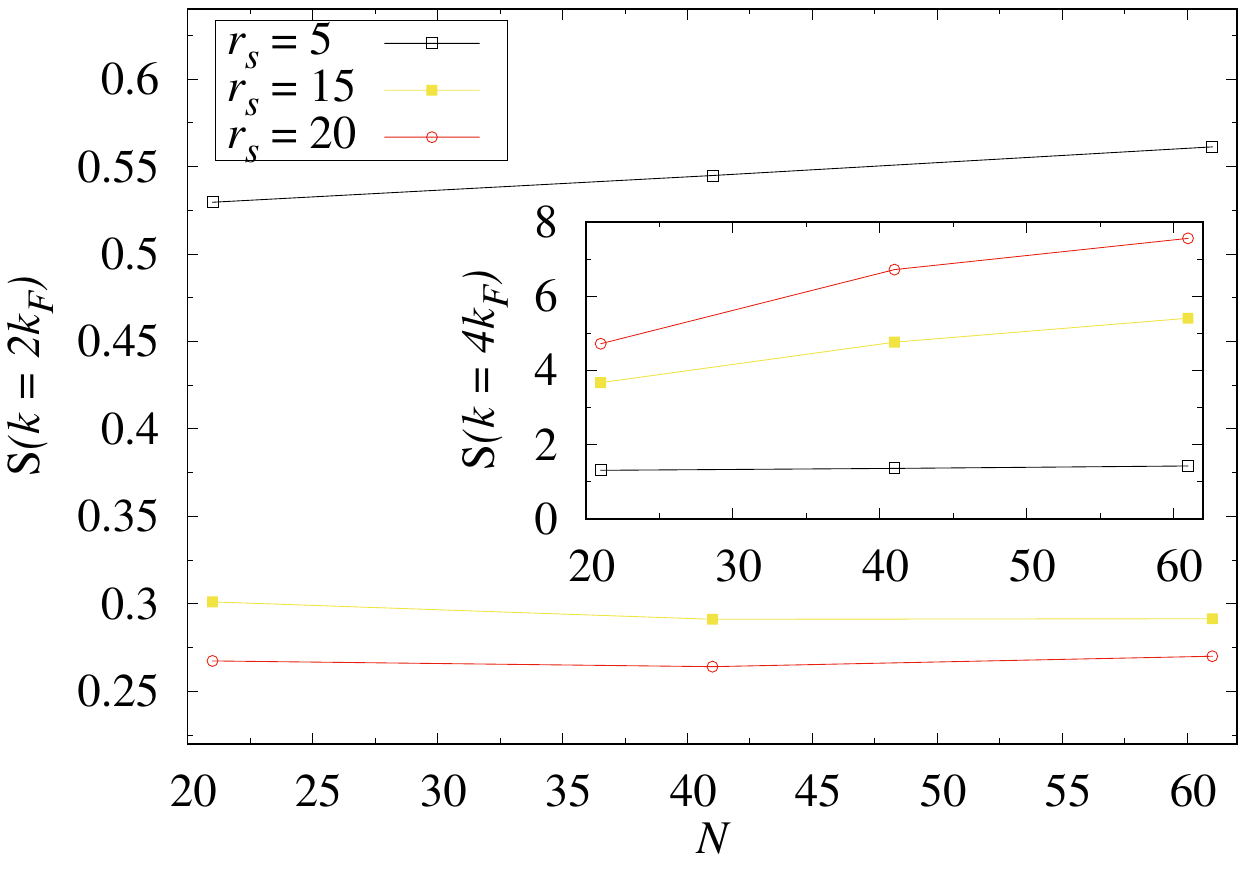}
\caption{\label{fig:SSF-N2} Peaks of SSF at different system sizes at $d = 1$ a.u.\ for $r_\text{s} = 5$, 15, and 20.}
\end{figure}

Figure \ref{fig:SSF-N} shows the SSF as a function of system size at $d=1$ a.u.\ for $r_\text{s}=2$ in the top panel and $r_\text{s}=5$ in the bottom. The inset in the top panel shows a zoomed-in view near $2k_\text{F}$, where one can see that the heights of the peaks corresponding to $N=41$ and $61$ are almost the same. However, the height of the peak at $k=4k_\text{F}$ (see bottom panel) is found to be relatively more sensitive to $N$; it increases sublinearly with $N$. Figure \ref{fig:SSF-N2} shows the height of peaks at $2k_\text{F}$ (main plot) and $4k_\text{F}$ (inset) as a function of system size at $d=1$ a.u.\ for $r_\text{s}=5$, $15$, and $20$. 

\begin{figure}[!htbp]
\centering
\includegraphics[width=.48\textwidth]{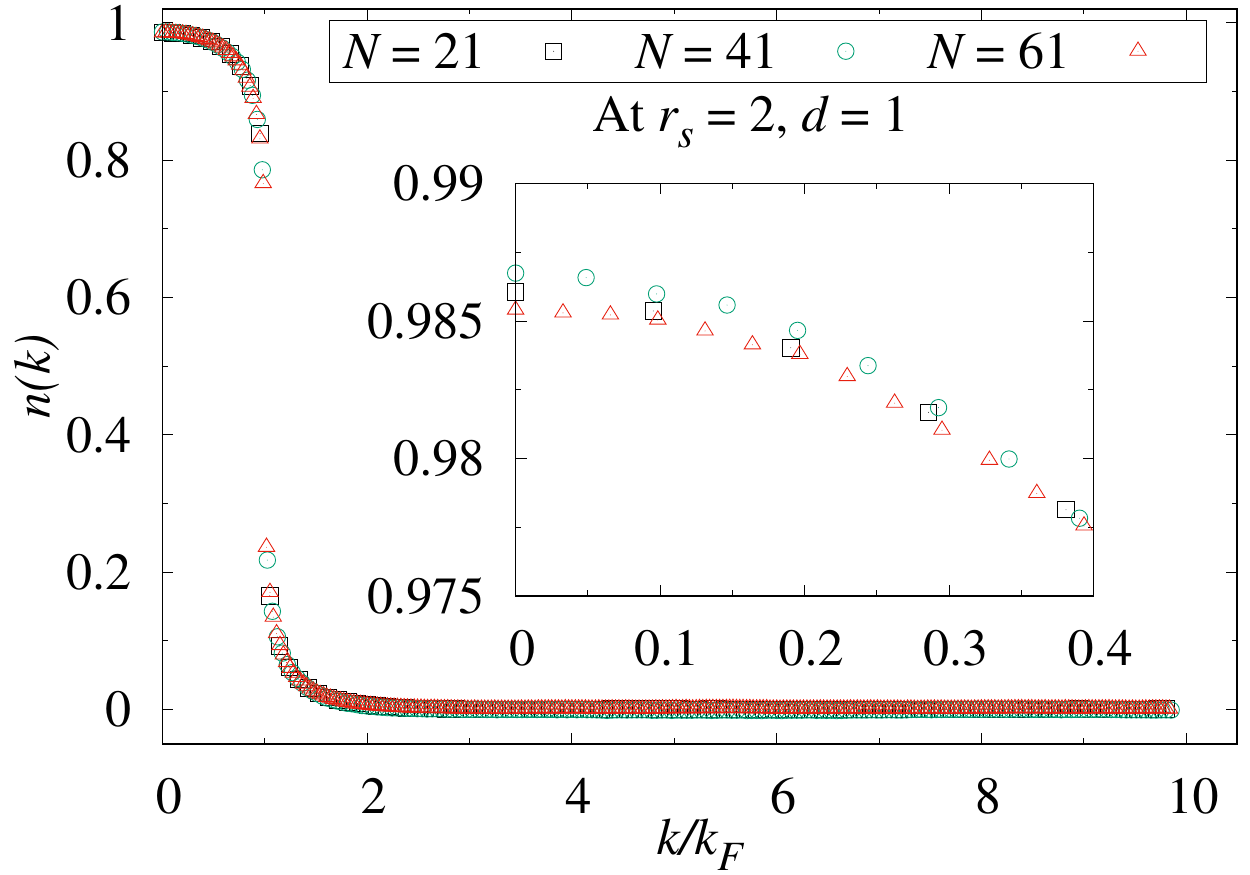}
\caption{\label{fig:MD-N} MD as a function of system size for $r_\text{s} = 2$ a.u.\ and $d = 1$ a.u. The inset shows a zoomed-in view for small $k$.}
\end{figure}

Figure \ref{fig:MD-N} shows the MD as a function of system size at $d=1$ a.u.\ for $r_\text{s} = 2$. The inset graph shows a zoomed-in view for small $k$, where one can see that the value of $n(k=0)$ slowly decreases with $N$. The finite-size effect on $n(k)$ is small.

\section{Comparison of VMC and DMC}\label{sec:V_DMC}

\begin{figure}[!htbp]
\centering
\includegraphics[width=.48\textwidth]{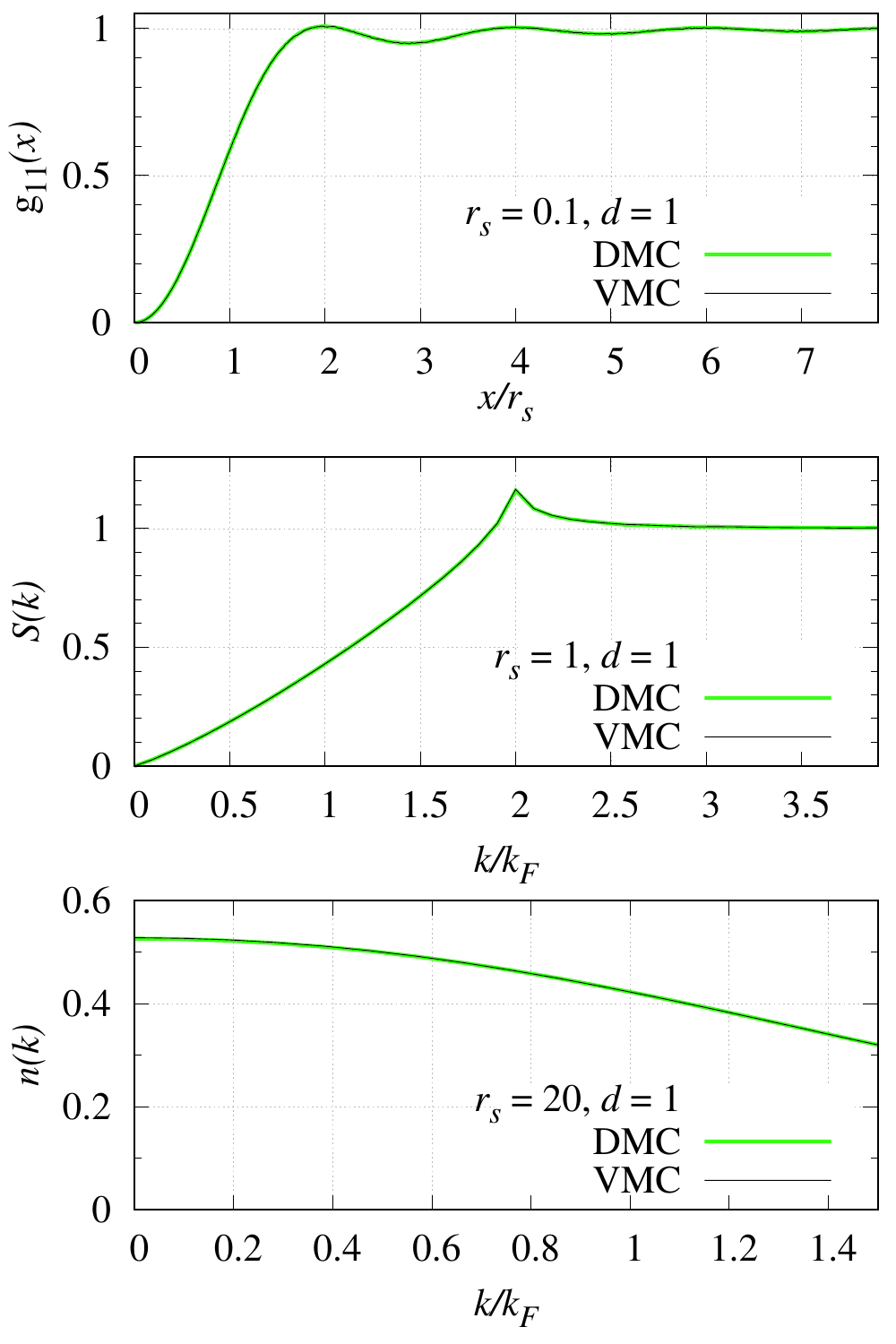}
\caption{\label{fig:vmc-dmc} Comparison of VMC and DMC PCFs, SSFs, and MDs at $d = 1$ a.u.\ for some values of $r_\text{s}$.}
\end{figure}

In this section we present DMC calculations performed to verify that VMC is sufficiently accurate in studies of EEBW systems. We choose a system size with $N=21$ and $d = 1$ a.u.\ at a few $r_\text{s}$ values for our DMC calculations. Table \ref{tab:dmc-vmc} shows the ground-state energy values computed using the VMC and DMC methods at $r_\text{s}=0.1$, 1, and 20. One can see that the VMC retries 99.98\% of correlation energy $E_\text{c}$. Comparisons of PCFs, SSFs, and MDs are shown in Fig.\ \ref{fig:vmc-dmc}. It is observed that the VMC and DMC values of these observables overlap, indicating that VMC is accurate enough for EEBW systems.

\begin{table}[!htbp]
\caption{\label{tab:dmc-vmc}Ground-state energy of the EEBW computed using VMC and DMC for $N=21$ and $d=1$ a.u.\ for various $r_\text{s}$. Percentage (\%) of correlation energy $E_{\text{c}}$, i.e., $(E_\text{VMC}-E_\text{HF})/(E_\text{DMC}-E_\text{HF})$ is calculated.}
\begin{ruledtabular}
\begin{tabular}{lcccc}
$r_\text{s}$ & VMC & DMC & $E_{\text{HF}}$& \% of $E_\text{C}$\\ \hline
0.1  &  50.151344(1)  &   50.151343(2)  &   50.280268  & 99.999224(3)\\
1.0  &   0.1502710(5) &    0.150255(4)  &    0.1756327 & 99.936953(4)\\ 
20.0 &$-$0.1429959(1) & $-$0.143006(3)  & $-$0.0856453 & 99.982392(3)
\end{tabular}
\end{ruledtabular}
\end{table}%

\section{Extrapolation of ground-state energies to the thermodynamic limit}\label{ap:chi2}

The energy data shown in Fig.\ \ref{fig:e-en} were extrapolated to infinite system size by fitting a model of the finite-size dependence to the data. We report here the reduced $\chi^2$ for various values of $r_\text{s}$ at $d = 1$ a.u.\ for the formulas $E(N)=E_\infty+B/N^2$ [Eq.\ (\ref{eqn:Ethrmdnm})], $E(N)=E_\infty+C\sqrt{\ln(N)}/N^2$, and $E(N)=E_\infty+C\sqrt{\ln(N)}/N^2+B/N^2$ \cite{Shulenburger2009}. Note that our simulation data for the ground state energy are only available for $N = 21$, 41, and 61; thus we do not expect the logarithmic term to make a significant difference, and we cannot assess the quality of the fit of the three-parameter model. From Table \ref{tab:chi-square} it is observed that the $1/N^2$ fit gives smaller reduced $\chi^2$ values than the $\sqrt{\ln(N)}/N^2$ fit for $r_\text{s} \leq 1$.  For larger $r_\text{s}$ the reduced $\chi^2$ values are similar. In all cases the extrapolated energies $E_\infty$ are almost the same.  The reduced $\chi^2$ values were calculated using the VMC error bars, and they are significantly larger than $1$, indicating that there are other sources of uncertainty in the VMC energy data $E(N)$.  These other sources of randomness in the data include the independent stochastic optimizations of the wave functions at different system sizes and quasi-random finite-size effects due to PCF oscillations being forced to be commensurate with the simulation cell.

\begin{table*}[!htbp]
\caption{\label{tab:chi-square} Extrapolated infinite-system energy per particle and reduced $\chi^2$ value obtained by fitting Eq.\ (\ref{eqn:Ethrmdnm}), $E(N)=E_\infty+C\ln(N)/N^2$, and $E(N)=E_\infty+C\ln(N)/N^2+B/N^2$ to VMC energy data at $d=1$ a.u.\ for various values of $r_\text{s}$. The VMC energy data are available at $N = 21$, 41, and 61.}
\begin{ruledtabular}
\begin{tabular}{lccccc}

& \multicolumn{3}{c}{Extrapolated energy $E_\infty$ (a.u./electron)} & \multicolumn{2}{c}{Reduced $\chi^2$ value} \\

\raisebox{1ex}[0pt]{$r_\text{s}$} & Eq.\ (\ref{eqn:Ethrmdnm}) & $E_\infty+C\sqrt{\ln(N)}/N^2$ & $E_\infty+C\sqrt{\ln(N)}/N^2+B/N^2$ & Eq.\ (\ref{eqn:Ethrmdnm}) & $E_\infty+C\sqrt{\ln(N)}/N^2$ \\ \hline

$0.1$ & $50.25337(7)$   & $50.2561(7)$    & $50.2536345116$ & $5.100 \times 10^{3}$ & $4.220 \times 10^{5}$ \\
$0.2$ & $13.10028(3)$   & $13.1012(2)$    & $13.1004336765$ & $4.286 \times 10^{3}$ & $1.089 \times 10^{5}$ \\
$0.4$ & $3.10151(1)$    & $3.10173(5)$    & $3.1015638274$  & $4.398 \times 10^{3}$ & $4.587 \times 10^{4}$ \\
$0.8$ & $0.463572(3)$   & $0.463627(9)$   & $0.4635877112$  & $3.054 \times 10^{2}$ & $1.997 \times 10^{3}$ \\
$1.0$ & $0.151823(3)$   & $0.151860(6)$   & $0.1518353048$  & $1.260 \times 10^{2}$ & $4.754 \times 10^{2}$ \\
$2.0$ & $-0.215851(2)$  & $-0.2158385(6)$ & $-0.2158409201$ & $1.163 \times 10^{1}$ & $7.300 \times 10^{-1}$ \\ 
$5.0$ & $-0.250173(2)$  & $-0.250168(1)$  & $-0.2501646650$ & $6.479$               & $1.275$                \\
$10.0$ & $-0.2048(5)$   & $-0.2048(5)$    & $-0.2028376433$ & $1.686 \times 10^{6}$ & $1.648 \times 10^{6}$ \\
$15.0$ & $-0.168301(7)$ & $-0.168294(5)$  & $-0.1682743481$ & $4.259 \times 10^{3}$ & $2.264 \times 10^{3}$ \\
$20.0$ & $-0.142888(6)$ & $-0.142885(5)$  & $-0.1428675352$ & $4.981 \times 10^{3}$ & $3.625 \times 10^{3}$
\end{tabular}
\end{ruledtabular}
\end{table*}%


\bibliography{library}

\begin{thebibliography}{72}%
\makeatletter
\providecommand \@ifxundefined [1]{%
 \@ifx{#1\undefined}
}%
\providecommand \@ifnum [1]{%
 \ifnum #1\expandafter \@firstoftwo
 \else \expandafter \@secondoftwo
 \fi
}%
\providecommand \@ifx [1]{%
 \ifx #1\expandafter \@firstoftwo
 \else \expandafter \@secondoftwo
 \fi
}%
\providecommand \natexlab [1]{#1}%
\providecommand \enquote  [1]{``#1''}%
\providecommand \bibnamefont  [1]{#1}%
\providecommand \bibfnamefont [1]{#1}%
\providecommand \citenamefont [1]{#1}%
\providecommand \href@noop [0]{\@secondoftwo}%
\providecommand \href [0]{\begingroup \@sanitize@url \@href}%
\providecommand \@href[1]{\@@startlink{#1}\@@href}%
\providecommand \@@href[1]{\endgroup#1\@@endlink}%
\providecommand \@sanitize@url [0]{\catcode `\\12\catcode `\$12\catcode
  `\&12\catcode `\#12\catcode `\^12\catcode `\_12\catcode `\%12\relax}%
\providecommand \@@startlink[1]{}%
\providecommand \@@endlink[0]{}%
\providecommand \url  [0]{\begingroup\@sanitize@url \@url }%
\providecommand \@url [1]{\endgroup\@href {#1}{\urlprefix }}%
\providecommand \urlprefix  [0]{URL }%
\providecommand \Eprint [0]{\href }%
\providecommand \doibase [0]{https://doi.org/}%
\providecommand \selectlanguage [0]{\@gobble}%
\providecommand \bibinfo  [0]{\@secondoftwo}%
\providecommand \bibfield  [0]{\@secondoftwo}%
\providecommand \translation [1]{[#1]}%
\providecommand \BibitemOpen [0]{}%
\providecommand \bibitemStop [0]{}%
\providecommand \bibitemNoStop [0]{.\EOS\space}%
\providecommand \EOS [0]{\spacefactor3000\relax}%
\providecommand \BibitemShut  [1]{\csname bibitem#1\endcsname}%
\let\auto@bib@innerbib\@empty
\bibitem [{\citenamefont {Go{\~{n}}i}\ \emph {et~al.}(1991)\citenamefont
  {Go{\~{n}}i}, \citenamefont {Pinczuk}, \citenamefont {Weiner}, \citenamefont
  {Calleja}, \citenamefont {Dennis}, \citenamefont {Pfeiffer},\ and\
  \citenamefont {West}}]{Goni1991}%
  \BibitemOpen
  \bibfield  {author} {\bibinfo {author} {\bibfnamefont {A.~R.}\ \bibnamefont
  {Go{\~{n}}i}}, \bibinfo {author} {\bibfnamefont {A.}~\bibnamefont {Pinczuk}},
  \bibinfo {author} {\bibfnamefont {J.~S.}\ \bibnamefont {Weiner}}, \bibinfo
  {author} {\bibfnamefont {J.~M.}\ \bibnamefont {Calleja}}, \bibinfo {author}
  {\bibfnamefont {B.~S.}\ \bibnamefont {Dennis}}, \bibinfo {author}
  {\bibfnamefont {L.~N.}\ \bibnamefont {Pfeiffer}},\ and\ \bibinfo {author}
  {\bibfnamefont {K.~W.}\ \bibnamefont {West}},\ }\href
  {https://doi.org/10.1103/PhysRevLett.67.3298} {\bibfield  {journal} {\bibinfo
   {journal} {Phys. Rev. Lett.}\ }\textbf {\bibinfo {volume} {67}},\ \bibinfo
  {pages} {3298} (\bibinfo {year} {1991})}\BibitemShut {NoStop}%
\bibitem [{\citenamefont {Altmann}\ \emph {et~al.}(2001)\citenamefont
  {Altmann}, \citenamefont {Crain}, \citenamefont {Kirakosian}, \citenamefont
  {Lin}, \citenamefont {Petrovykh}, \citenamefont {Himpsel},\ and\
  \citenamefont {Losio}}]{Altmann2001}%
  \BibitemOpen
  \bibfield  {author} {\bibinfo {author} {\bibfnamefont {K.~N.}\ \bibnamefont
  {Altmann}}, \bibinfo {author} {\bibfnamefont {J.~N.}\ \bibnamefont {Crain}},
  \bibinfo {author} {\bibfnamefont {A.}~\bibnamefont {Kirakosian}}, \bibinfo
  {author} {\bibfnamefont {J.-L.}\ \bibnamefont {Lin}}, \bibinfo {author}
  {\bibfnamefont {D.~Y.}\ \bibnamefont {Petrovykh}}, \bibinfo {author}
  {\bibfnamefont {F.~J.}\ \bibnamefont {Himpsel}},\ and\ \bibinfo {author}
  {\bibfnamefont {R.}~\bibnamefont {Losio}},\ }\href
  {https://doi.org/10.1103/PhysRevB.64.035406} {\bibfield  {journal} {\bibinfo
  {journal} {Phys. Rev. B}\ }\textbf {\bibinfo {volume} {64}},\ \bibinfo
  {pages} {035406} (\bibinfo {year} {2001})}\BibitemShut {NoStop}%
\bibitem [{\citenamefont {Nagao}\ \emph {et~al.}(2006)\citenamefont {Nagao},
  \citenamefont {Yaginuma}, \citenamefont {Inaoka},\ and\ \citenamefont
  {Sakurai}}]{Nagao2006}%
  \BibitemOpen
  \bibfield  {author} {\bibinfo {author} {\bibfnamefont {T.}~\bibnamefont
  {Nagao}}, \bibinfo {author} {\bibfnamefont {S.}~\bibnamefont {Yaginuma}},
  \bibinfo {author} {\bibfnamefont {T.}~\bibnamefont {Inaoka}},\ and\ \bibinfo
  {author} {\bibfnamefont {T.}~\bibnamefont {Sakurai}},\ }\href
  {https://doi.org/10.1103/PhysRevLett.97.116802} {\bibfield  {journal}
  {\bibinfo  {journal} {Phys. Rev. Lett.}\ }\textbf {\bibinfo {volume} {97}},\
  \bibinfo {pages} {116802} (\bibinfo {year} {2006})}\BibitemShut {NoStop}%
\bibitem [{\citenamefont {Hong}\ \emph {et~al.}(2016)\citenamefont {Hong},
  \citenamefont {Zhang}, \citenamefont {Zhang}, \citenamefont {Zhang},
  \citenamefont {Wang}, \citenamefont {Chen}, \citenamefont {Shen},\ and\
  \citenamefont {Sun}}]{Hong2016}%
  \BibitemOpen
  \bibfield  {author} {\bibinfo {author} {\bibfnamefont {D.~S.}\ \bibnamefont
  {Hong}}, \bibinfo {author} {\bibfnamefont {H.}~\bibnamefont {Zhang}},
  \bibinfo {author} {\bibfnamefont {H.~R.}\ \bibnamefont {Zhang}}, \bibinfo
  {author} {\bibfnamefont {J.}~\bibnamefont {Zhang}}, \bibinfo {author}
  {\bibfnamefont {S.~F.}\ \bibnamefont {Wang}}, \bibinfo {author}
  {\bibfnamefont {Y.~S.}\ \bibnamefont {Chen}}, \bibinfo {author}
  {\bibfnamefont {B.~G.}\ \bibnamefont {Shen}},\ and\ \bibinfo {author}
  {\bibfnamefont {J.~R.}\ \bibnamefont {Sun}},\ }\href
  {https://doi.org/10.1063/1.4966546} {\bibfield  {journal} {\bibinfo
  {journal} {Appl. Phys. Lett.}\ }\textbf {\bibinfo {volume} {109}},\ \bibinfo
  {pages} {173505} (\bibinfo {year} {2016})}\BibitemShut {NoStop}%
\bibitem [{\citenamefont {Friesen}\ and\ \citenamefont
  {Bergersen}(1980)}]{Friesen1980}%
  \BibitemOpen
  \bibfield  {author} {\bibinfo {author} {\bibfnamefont {W.~I.}\ \bibnamefont
  {Friesen}}\ and\ \bibinfo {author} {\bibfnamefont {B.}~\bibnamefont
  {Bergersen}},\ }\href {https://doi.org/10.1088/0022-3719/13/36/016}
  {\bibfield  {journal} {\bibinfo  {journal} {J. Phys. C: Solid State Phys.}\
  }\textbf {\bibinfo {volume} {13}},\ \bibinfo {pages} {6627} (\bibinfo {year}
  {1980})}\BibitemShut {NoStop}%
\bibitem [{\citenamefont {{Das Sarma}}\ and\ \citenamefont
  {Lai}(1985)}]{DasSarma1985}%
  \BibitemOpen
  \bibfield  {author} {\bibinfo {author} {\bibfnamefont {S.}~\bibnamefont {{Das
  Sarma}}}\ and\ \bibinfo {author} {\bibfnamefont {W.}~\bibnamefont {Lai}},\
  }\href {https://doi.org/10.1103/PhysRevB.32.1401} {\bibfield  {journal}
  {\bibinfo  {journal} {Phys. Rev. B}\ }\textbf {\bibinfo {volume} {32}},\
  \bibinfo {pages} {1401} (\bibinfo {year} {1985})}\BibitemShut {NoStop}%
\bibitem [{\citenamefont {Schulz}(1993)}]{Schulz1993}%
  \BibitemOpen
  \bibfield  {author} {\bibinfo {author} {\bibfnamefont {H.~J.}\ \bibnamefont
  {Schulz}},\ }\href {https://doi.org/10.1103/PhysRevLett.71.1864} {\bibfield
  {journal} {\bibinfo  {journal} {Phys. Rev. Lett.}\ }\textbf {\bibinfo
  {volume} {71}},\ \bibinfo {pages} {1864} (\bibinfo {year}
  {1993})}\BibitemShut {NoStop}%
\bibitem [{\citenamefont {Tanatar}\ \emph {et~al.}(1998)\citenamefont
  {Tanatar}, \citenamefont {Al-Hayek},\ and\ \citenamefont
  {Tomak}}]{Tanatar1998c}%
  \BibitemOpen
  \bibfield  {author} {\bibinfo {author} {\bibfnamefont {B.}~\bibnamefont
  {Tanatar}}, \bibinfo {author} {\bibfnamefont {I.}~\bibnamefont {Al-Hayek}},\
  and\ \bibinfo {author} {\bibfnamefont {M.}~\bibnamefont {Tomak}},\ }\href
  {https://doi.org/10.1103/PhysRevB.58.9886} {\bibfield  {journal} {\bibinfo
  {journal} {Phys. Rev. B}\ }\textbf {\bibinfo {volume} {58}},\ \bibinfo
  {pages} {9886} (\bibinfo {year} {1998})}\BibitemShut {NoStop}%
\bibitem [{\citenamefont {Moudgil}\ \emph
  {et~al.}(2010{\natexlab{a}})\citenamefont {Moudgil}, \citenamefont {Garg},\
  and\ \citenamefont {Pathak}}]{Moudgil2010a}%
  \BibitemOpen
  \bibfield  {author} {\bibinfo {author} {\bibfnamefont {R.~K.}\ \bibnamefont
  {Moudgil}}, \bibinfo {author} {\bibfnamefont {V.}~\bibnamefont {Garg}},\ and\
  \bibinfo {author} {\bibfnamefont {K.~N.}\ \bibnamefont {Pathak}},\ }\href
  {https://doi.org/10.1088/0953-8984/22/13/135003} {\bibfield  {journal}
  {\bibinfo  {journal} {J. Phys.: Condens. Matter}\ }\textbf {\bibinfo {volume}
  {22}},\ \bibinfo {pages} {135003} (\bibinfo {year}
  {2010}{\natexlab{a}})}\BibitemShut {NoStop}%
\bibitem [{\citenamefont {Tomonaga}(1950)}]{Tomonaga1950}%
  \BibitemOpen
  \bibfield  {author} {\bibinfo {author} {\bibfnamefont {S.}~\bibnamefont
  {Tomonaga}},\ }\href {https://doi.org/10.1143/ptp/5.4.544} {\bibfield
  {journal} {\bibinfo  {journal} {Prog. Theor. Phys.}\ }\textbf {\bibinfo
  {volume} {5}},\ \bibinfo {pages} {544} (\bibinfo {year} {1950})}\BibitemShut
  {NoStop}%
\bibitem [{\citenamefont {Luttinger}(1963)}]{Luttinger1963}%
  \BibitemOpen
  \bibfield  {author} {\bibinfo {author} {\bibfnamefont {J.~M.}\ \bibnamefont
  {Luttinger}},\ }\href {https://doi.org/10.1063/1.1704046} {\bibfield
  {journal} {\bibinfo  {journal} {J. Math. Phys.}\ }\textbf {\bibinfo {volume}
  {4}},\ \bibinfo {pages} {1154} (\bibinfo {year} {1963})}\BibitemShut
  {NoStop}%
\bibitem [{\citenamefont {Haldane}(1981)}]{Haldane1982}%
  \BibitemOpen
  \bibfield  {author} {\bibinfo {author} {\bibfnamefont {F.~D.~M.}\
  \bibnamefont {Haldane}},\ }\href
  {https://doi.org/10.1103/PhysRevLett.47.1840} {\bibfield  {journal} {\bibinfo
   {journal} {Phys. Rev. Lett.}\ }\textbf {\bibinfo {volume} {47}},\ \bibinfo
  {pages} {1840} (\bibinfo {year} {1981})}\BibitemShut {NoStop}%
\bibitem [{\citenamefont {Giamarchi}(2003)}]{Giamarchi2003}%
  \BibitemOpen
  \bibfield  {author} {\bibinfo {author} {\bibfnamefont {T.}~\bibnamefont
  {Giamarchi}},\ }\href
  {https://doi.org/10.1093/acprof:oso/9780198525004.001.0001} {\emph {\bibinfo
  {title} {Quantum Physics in One Dimension}}}\ (\bibinfo  {publisher} {Oxford
  University Press},\ \bibinfo {year} {2003})\BibitemShut {NoStop}%
\bibitem [{\citenamefont {Bala}\ \emph {et~al.}(2014)\citenamefont {Bala},
  \citenamefont {Moudgil}, \citenamefont {Srivastava},\ and\ \citenamefont
  {Pathak}}]{Bala2014}%
  \BibitemOpen
  \bibfield  {author} {\bibinfo {author} {\bibfnamefont {R.}~\bibnamefont
  {Bala}}, \bibinfo {author} {\bibfnamefont {R.~K.}\ \bibnamefont {Moudgil}},
  \bibinfo {author} {\bibfnamefont {S.}~\bibnamefont {Srivastava}},\ and\
  \bibinfo {author} {\bibfnamefont {K.~N.}\ \bibnamefont {Pathak}},\ }\href
  {https://doi.org/10.1140/epjb/e2013-40567-3} {\bibfield  {journal} {\bibinfo
  {journal} {Eur. Phys. J. B}\ }\textbf {\bibinfo {volume} {87}},\ \bibinfo
  {pages} {5} (\bibinfo {year} {2014})}\BibitemShut {NoStop}%
\bibitem [{\citenamefont {Ashokan}\ \emph
  {et~al.}(2018{\natexlab{a}})\citenamefont {Ashokan}, \citenamefont {Bala},
  \citenamefont {Morawetz},\ and\ \citenamefont {Pathak}}]{Ashokan2018}%
  \BibitemOpen
  \bibfield  {author} {\bibinfo {author} {\bibfnamefont {V.}~\bibnamefont
  {Ashokan}}, \bibinfo {author} {\bibfnamefont {R.}~\bibnamefont {Bala}},
  \bibinfo {author} {\bibfnamefont {K.}~\bibnamefont {Morawetz}},\ and\
  \bibinfo {author} {\bibfnamefont {K.~N.}\ \bibnamefont {Pathak}},\ }\href
  {https://doi.org/10.1140/epjb/e2017-80530-8} {\bibfield  {journal} {\bibinfo
  {journal} {Eur. Phys. J. B}\ }\textbf {\bibinfo {volume} {91}},\ \bibinfo
  {pages} {29} (\bibinfo {year} {2018}{\natexlab{a}})}\BibitemShut {NoStop}%
\bibitem [{\citenamefont {Morawetz}\ \emph {et~al.}(2018)\citenamefont
  {Morawetz}, \citenamefont {Ashokan}, \citenamefont {Bala},\ and\
  \citenamefont {Pathak}}]{Morawetz2018}%
  \BibitemOpen
  \bibfield  {author} {\bibinfo {author} {\bibfnamefont {K.}~\bibnamefont
  {Morawetz}}, \bibinfo {author} {\bibfnamefont {V.}~\bibnamefont {Ashokan}},
  \bibinfo {author} {\bibfnamefont {R.}~\bibnamefont {Bala}},\ and\ \bibinfo
  {author} {\bibfnamefont {K.~N.}\ \bibnamefont {Pathak}},\ }\href
  {https://doi.org/10.1103/PhysRevB.97.155147} {\bibfield  {journal} {\bibinfo
  {journal} {Phys. Rev. B}\ }\textbf {\bibinfo {volume} {97}},\ \bibinfo
  {pages} {155147} (\bibinfo {year} {2018})}\BibitemShut {NoStop}%
\bibitem [{\citenamefont {Ashokan}\ \emph {et~al.}(2020)\citenamefont
  {Ashokan}, \citenamefont {Bala}, \citenamefont {Morawetz},\ and\
  \citenamefont {Pathak}}]{Ashokan2020}%
  \BibitemOpen
  \bibfield  {author} {\bibinfo {author} {\bibfnamefont {V.}~\bibnamefont
  {Ashokan}}, \bibinfo {author} {\bibfnamefont {R.}~\bibnamefont {Bala}},
  \bibinfo {author} {\bibfnamefont {K.}~\bibnamefont {Morawetz}},\ and\
  \bibinfo {author} {\bibfnamefont {K.~N.}\ \bibnamefont {Pathak}},\ }\href
  {https://doi.org/10.1103/PhysRevB.101.075130} {\bibfield  {journal} {\bibinfo
   {journal} {Phys. Rev. B}\ }\textbf {\bibinfo {volume} {101}},\ \bibinfo
  {pages} {075130} (\bibinfo {year} {2020})}\BibitemShut {NoStop}%
\bibitem [{\citenamefont {Tanatar}\ and\ \citenamefont
  {Bulutay}(1999)}]{Tanatar1999}%
  \BibitemOpen
  \bibfield  {author} {\bibinfo {author} {\bibfnamefont {B.}~\bibnamefont
  {Tanatar}}\ and\ \bibinfo {author} {\bibfnamefont {C.}~\bibnamefont
  {Bulutay}},\ }\href {https://doi.org/10.1103/PhysRevB.59.15019} {\bibfield
  {journal} {\bibinfo  {journal} {Phys. Rev. B}\ }\textbf {\bibinfo {volume}
  {59}},\ \bibinfo {pages} {15019} (\bibinfo {year} {1999})}\BibitemShut
  {NoStop}%
\bibitem [{\citenamefont {Demirel}\ and\ \citenamefont
  {Tanatar}(1999)}]{Demirel1999a}%
  \BibitemOpen
  \bibfield  {author} {\bibinfo {author} {\bibfnamefont {E.}~\bibnamefont
  {Demirel}}\ and\ \bibinfo {author} {\bibfnamefont {B.}~\bibnamefont
  {Tanatar}},\ }\href {https://doi.org/10.1007/s100510050975} {\bibfield
  {journal} {\bibinfo  {journal} {Eur. Phys. J. B}\ }\textbf {\bibinfo {volume}
  {12}},\ \bibinfo {pages} {47} (\bibinfo {year} {1999})}\BibitemShut {NoStop}%
\bibitem [{\citenamefont {Garg}\ \emph {et~al.}(2008)\citenamefont {Garg},
  \citenamefont {Moudgil}, \citenamefont {Kumar},\ and\ \citenamefont
  {Ahluwalia}}]{Garg2008}%
  \BibitemOpen
  \bibfield  {author} {\bibinfo {author} {\bibfnamefont {V.}~\bibnamefont
  {Garg}}, \bibinfo {author} {\bibfnamefont {R.~K.}\ \bibnamefont {Moudgil}},
  \bibinfo {author} {\bibfnamefont {K.}~\bibnamefont {Kumar}},\ and\ \bibinfo
  {author} {\bibfnamefont {P.~K.}\ \bibnamefont {Ahluwalia}},\ }\href
  {https://doi.org/10.1103/PhysRevB.78.045406} {\bibfield  {journal} {\bibinfo
  {journal} {Phys. Rev. B}\ }\textbf {\bibinfo {volume} {78}},\ \bibinfo
  {pages} {045406} (\bibinfo {year} {2008})}\BibitemShut {NoStop}%
\bibitem [{\citenamefont {Sharma}\ \emph
  {et~al.}(2018{\natexlab{a}})\citenamefont {Sharma}, \citenamefont {Kaur},
  \citenamefont {Garg},\ and\ \citenamefont {Moudgil}}]{Sharma2018c}%
  \BibitemOpen
  \bibfield  {author} {\bibinfo {author} {\bibfnamefont {A.}~\bibnamefont
  {Sharma}}, \bibinfo {author} {\bibfnamefont {K.}~\bibnamefont {Kaur}},
  \bibinfo {author} {\bibfnamefont {V.}~\bibnamefont {Garg}},\ and\ \bibinfo
  {author} {\bibfnamefont {R.~K.}\ \bibnamefont {Moudgil}},\ }\href
  {https://doi.org/10.1002/pssb.201800174} {\bibfield  {journal} {\bibinfo
  {journal} {Phys. Status Solidi B}\ }\textbf {\bibinfo {volume} {255}},\
  \bibinfo {pages} {1800174} (\bibinfo {year}
  {2018}{\natexlab{a}})}\BibitemShut {NoStop}%
\bibitem [{\citenamefont {Casula}\ \emph {et~al.}(2006)\citenamefont {Casula},
  \citenamefont {Sorella},\ and\ \citenamefont {Senatore}}]{Casula2006}%
  \BibitemOpen
  \bibfield  {author} {\bibinfo {author} {\bibfnamefont {M.}~\bibnamefont
  {Casula}}, \bibinfo {author} {\bibfnamefont {S.}~\bibnamefont {Sorella}},\
  and\ \bibinfo {author} {\bibfnamefont {G.}~\bibnamefont {Senatore}},\ }\href
  {https://doi.org/10.1103/PhysRevB.74.245427} {\bibfield  {journal} {\bibinfo
  {journal} {Phys. Rev. B}\ }\textbf {\bibinfo {volume} {74}},\ \bibinfo
  {pages} {245427} (\bibinfo {year} {2006})}\BibitemShut {NoStop}%
\bibitem [{\citenamefont {Shulenburger}\ \emph {et~al.}(2008)\citenamefont
  {Shulenburger}, \citenamefont {Casula}, \citenamefont {Senatore},\ and\
  \citenamefont {Martin}}]{Shulenburger2008}%
  \BibitemOpen
  \bibfield  {author} {\bibinfo {author} {\bibfnamefont {L.}~\bibnamefont
  {Shulenburger}}, \bibinfo {author} {\bibfnamefont {M.}~\bibnamefont
  {Casula}}, \bibinfo {author} {\bibfnamefont {G.}~\bibnamefont {Senatore}},\
  and\ \bibinfo {author} {\bibfnamefont {R.~M.}\ \bibnamefont {Martin}},\
  }\href {https://doi.org/10.1103/PhysRevB.78.165303} {\bibfield  {journal}
  {\bibinfo  {journal} {Phys. Rev. B}\ }\textbf {\bibinfo {volume} {78}},\
  \bibinfo {pages} {165303} (\bibinfo {year} {2008})}\BibitemShut {NoStop}%
\bibitem [{\citenamefont {Lee}\ and\ \citenamefont
  {Drummond}(2011)}]{Lee2011b}%
  \BibitemOpen
  \bibfield  {author} {\bibinfo {author} {\bibfnamefont {R.~M.}\ \bibnamefont
  {Lee}}\ and\ \bibinfo {author} {\bibfnamefont {N.~D.}\ \bibnamefont
  {Drummond}},\ }\href {https://doi.org/10.1103/PhysRevB.83.245114} {\bibfield
  {journal} {\bibinfo  {journal} {Phys. Rev. B}\ }\textbf {\bibinfo {volume}
  {83}},\ \bibinfo {pages} {245114} (\bibinfo {year} {2011})}\BibitemShut
  {NoStop}%
\bibitem [{\citenamefont {Ashokan}\ \emph
  {et~al.}(2018{\natexlab{b}})\citenamefont {Ashokan}, \citenamefont
  {Drummond},\ and\ \citenamefont {Pathak}}]{Ashokan2018a}%
  \BibitemOpen
  \bibfield  {author} {\bibinfo {author} {\bibfnamefont {V.}~\bibnamefont
  {Ashokan}}, \bibinfo {author} {\bibfnamefont {N.~D.}\ \bibnamefont
  {Drummond}},\ and\ \bibinfo {author} {\bibfnamefont {K.~N.}\ \bibnamefont
  {Pathak}},\ }\href {https://doi.org/10.1103/PhysRevB.98.125139} {\bibfield
  {journal} {\bibinfo  {journal} {Phys. Rev. B}\ }\textbf {\bibinfo {volume}
  {98}},\ \bibinfo {pages} {125139} (\bibinfo {year}
  {2018}{\natexlab{b}})}\BibitemShut {NoStop}%
\bibitem [{\citenamefont {Senatore}\ and\ \citenamefont {{De
  Palo}}(2003)}]{Senatore2003}%
  \BibitemOpen
  \bibfield  {author} {\bibinfo {author} {\bibfnamefont {G.}~\bibnamefont
  {Senatore}}\ and\ \bibinfo {author} {\bibfnamefont {S.}~\bibnamefont {{De
  Palo}}},\ }\href {https://doi.org/10.1002/ctpp.200310047} {\bibfield
  {journal} {\bibinfo  {journal} {Contrib. Plasma Phys.}\ }\textbf {\bibinfo
  {volume} {43}},\ \bibinfo {pages} {363} (\bibinfo {year} {2003})}\BibitemShut
  {NoStop}%
\bibitem [{\citenamefont {Kou}\ \emph {et~al.}(2014)\citenamefont {Kou},
  \citenamefont {Feldman}, \citenamefont {Levin}, \citenamefont {Halperin},
  \citenamefont {Watanabe}, \citenamefont {Taniguchi},\ and\ \citenamefont
  {Yacoby}}]{Kou2014}%
  \BibitemOpen
  \bibfield  {author} {\bibinfo {author} {\bibfnamefont {A.}~\bibnamefont
  {Kou}}, \bibinfo {author} {\bibfnamefont {B.~E.}\ \bibnamefont {Feldman}},
  \bibinfo {author} {\bibfnamefont {A.~J.}\ \bibnamefont {Levin}}, \bibinfo
  {author} {\bibfnamefont {B.~I.}\ \bibnamefont {Halperin}}, \bibinfo {author}
  {\bibfnamefont {K.}~\bibnamefont {Watanabe}}, \bibinfo {author}
  {\bibfnamefont {T.}~\bibnamefont {Taniguchi}},\ and\ \bibinfo {author}
  {\bibfnamefont {A.}~\bibnamefont {Yacoby}},\ }\href
  {https://doi.org/10.1126/science.1250270} {\bibfield  {journal} {\bibinfo
  {journal} {Science}\ }\textbf {\bibinfo {volume} {345}},\ \bibinfo {pages}
  {55} (\bibinfo {year} {2014})}\BibitemShut {NoStop}%
\bibitem [{\citenamefont {Sharma}\ \emph {et~al.}(2016)\citenamefont {Sharma},
  \citenamefont {Saini},\ and\ \citenamefont {Bahuguna}}]{Sharma2016}%
  \BibitemOpen
  \bibfield  {author} {\bibinfo {author} {\bibfnamefont {R.~O.}\ \bibnamefont
  {Sharma}}, \bibinfo {author} {\bibfnamefont {L.~K.}\ \bibnamefont {Saini}},\
  and\ \bibinfo {author} {\bibfnamefont {B.~P.}\ \bibnamefont {Bahuguna}},\
  }\href {https://doi.org/10.1103/PhysRevB.94.205435} {\bibfield  {journal}
  {\bibinfo  {journal} {Phys. Rev. B}\ }\textbf {\bibinfo {volume} {94}},\
  \bibinfo {pages} {205435} (\bibinfo {year} {2016})}\BibitemShut {NoStop}%
\bibitem [{\citenamefont {Butov}(2017)}]{Butov2017}%
  \BibitemOpen
  \bibfield  {author} {\bibinfo {author} {\bibfnamefont {L.}~\bibnamefont
  {Butov}},\ }\href {https://doi.org/10.1016/j.spmi.2016.12.035} {\bibfield
  {journal} {\bibinfo  {journal} {Superlattice Microst.}\ }\textbf {\bibinfo
  {volume} {108}},\ \bibinfo {pages} {2} (\bibinfo {year} {2017})}\BibitemShut
  {NoStop}%
\bibitem [{\citenamefont {Sharma}\ \emph {et~al.}(2017)\citenamefont {Sharma},
  \citenamefont {Saini},\ and\ \citenamefont {Bahuguna}}]{Sharma2017}%
  \BibitemOpen
  \bibfield  {author} {\bibinfo {author} {\bibfnamefont {R.~O.}\ \bibnamefont
  {Sharma}}, \bibinfo {author} {\bibfnamefont {L.~K.}\ \bibnamefont {Saini}},\
  and\ \bibinfo {author} {\bibfnamefont {B.~P.}\ \bibnamefont {Bahuguna}},\
  }\href {https://doi.org/10.1039/C7CP02934A} {\bibfield  {journal} {\bibinfo
  {journal} {Phys. Chem. Chem. Phys.}\ }\textbf {\bibinfo {volume} {19}},\
  \bibinfo {pages} {20778} (\bibinfo {year} {2017})}\BibitemShut {NoStop}%
\bibitem [{\citenamefont {{L{\'{o}}pez R{\'{i}}os}}\ \emph
  {et~al.}(2018)\citenamefont {{L{\'{o}}pez R{\'{i}}os}}, \citenamefont
  {Perali}, \citenamefont {Needs},\ and\ \citenamefont
  {Neilson}}]{LopezRios2018}%
  \BibitemOpen
  \bibfield  {author} {\bibinfo {author} {\bibfnamefont {P.}~\bibnamefont
  {{L{\'{o}}pez R{\'{i}}os}}}, \bibinfo {author} {\bibfnamefont
  {A.}~\bibnamefont {Perali}}, \bibinfo {author} {\bibfnamefont {R.~J.}\
  \bibnamefont {Needs}},\ and\ \bibinfo {author} {\bibfnamefont
  {D.}~\bibnamefont {Neilson}},\ }\href
  {https://doi.org/10.1103/PhysRevLett.120.177701} {\bibfield  {journal}
  {\bibinfo  {journal} {Phys. Rev. Lett.}\ }\textbf {\bibinfo {volume} {120}},\
  \bibinfo {pages} {177701} (\bibinfo {year} {2018})}\BibitemShut {NoStop}%
\bibitem [{\citenamefont {Sharma}\ \emph
  {et~al.}(2018{\natexlab{b}})\citenamefont {Sharma}, \citenamefont {Saini},\
  and\ \citenamefont {Bahuguna}}]{Sharma2018}%
  \BibitemOpen
  \bibfield  {author} {\bibinfo {author} {\bibfnamefont {R.~O.}\ \bibnamefont
  {Sharma}}, \bibinfo {author} {\bibfnamefont {L.~K.}\ \bibnamefont {Saini}},\
  and\ \bibinfo {author} {\bibfnamefont {B.~P.}\ \bibnamefont {Bahuguna}},\
  }\href {https://doi.org/10.1088/1361-648X/aab81c} {\bibfield  {journal}
  {\bibinfo  {journal} {J. Phys.: Condens. Matter}\ }\textbf {\bibinfo {volume}
  {30}},\ \bibinfo {pages} {185404} (\bibinfo {year}
  {2018}{\natexlab{b}})}\BibitemShut {NoStop}%
\bibitem [{\citenamefont {Yang}\ \emph {et~al.}(2020)\citenamefont {Yang},
  \citenamefont {Perrin}, \citenamefont {Petrescu}, \citenamefont {Garate},\
  and\ \citenamefont {{Le Hur}}}]{Yang2020}%
  \BibitemOpen
  \bibfield  {author} {\bibinfo {author} {\bibfnamefont {F.}~\bibnamefont
  {Yang}}, \bibinfo {author} {\bibfnamefont {V.}~\bibnamefont {Perrin}},
  \bibinfo {author} {\bibfnamefont {A.}~\bibnamefont {Petrescu}}, \bibinfo
  {author} {\bibfnamefont {I.}~\bibnamefont {Garate}},\ and\ \bibinfo {author}
  {\bibfnamefont {K.}~\bibnamefont {{Le Hur}}},\ }\href
  {https://doi.org/10.1103/PhysRevB.101.085116} {\bibfield  {journal} {\bibinfo
   {journal} {Phys. Rev. B}\ }\textbf {\bibinfo {volume} {101}},\ \bibinfo
  {pages} {085116} (\bibinfo {year} {2020})}\BibitemShut {NoStop}%
\bibitem [{\citenamefont {Li}\ \emph {et~al.}(2020)\citenamefont {Li},
  \citenamefont {Ebisu}, \citenamefont {Sahoo}, \citenamefont {Oreg},\ and\
  \citenamefont {Franz}}]{Li2020}%
  \BibitemOpen
  \bibfield  {author} {\bibinfo {author} {\bibfnamefont {C.}~\bibnamefont
  {Li}}, \bibinfo {author} {\bibfnamefont {H.}~\bibnamefont {Ebisu}}, \bibinfo
  {author} {\bibfnamefont {S.}~\bibnamefont {Sahoo}}, \bibinfo {author}
  {\bibfnamefont {Y.}~\bibnamefont {Oreg}},\ and\ \bibinfo {author}
  {\bibfnamefont {M.}~\bibnamefont {Franz}},\ }\href
  {https://link.aps.org/doi/10.1103/PhysRevB.102.165123} {\bibfield  {journal}
  {\bibinfo  {journal} {Phys. Rev. B}\ }\textbf {\bibinfo {volume} {102}},\
  \bibinfo {pages} {165123} (\bibinfo {year} {2020})}\BibitemShut {NoStop}%
\bibitem [{\citenamefont {Meng}(2020)}]{Meng2019}%
  \BibitemOpen
  \bibfield  {author} {\bibinfo {author} {\bibfnamefont {T.}~\bibnamefont
  {Meng}},\ }\href {https://doi.org/10.1140/epjst/e2019-900095-5} {\bibfield
  {journal} {\bibinfo  {journal} {Eur. Phys. J. Spec. Top.}\ }\textbf {\bibinfo
  {volume} {229}},\ \bibinfo {pages} {527} (\bibinfo {year}
  {2020})}\BibitemShut {NoStop}%
\bibitem [{\citenamefont {Fuji}\ and\ \citenamefont
  {Furusaki}(2019)}]{Fuji2019}%
  \BibitemOpen
  \bibfield  {author} {\bibinfo {author} {\bibfnamefont {Y.}~\bibnamefont
  {Fuji}}\ and\ \bibinfo {author} {\bibfnamefont {A.}~\bibnamefont
  {Furusaki}},\ }\href {https://doi.org/10.1103/PhysRevB.99.035130} {\bibfield
  {journal} {\bibinfo  {journal} {Phys. Rev. B}\ }\textbf {\bibinfo {volume}
  {99}},\ \bibinfo {pages} {035130} (\bibinfo {year} {2019})}\BibitemShut
  {NoStop}%
\bibitem [{\citenamefont {Iadecola}\ \emph {et~al.}(2019)\citenamefont
  {Iadecola}, \citenamefont {Neupert}, \citenamefont {Chamon},\ and\
  \citenamefont {Mudry}}]{Iadecola2019}%
  \BibitemOpen
  \bibfield  {author} {\bibinfo {author} {\bibfnamefont {T.}~\bibnamefont
  {Iadecola}}, \bibinfo {author} {\bibfnamefont {T.}~\bibnamefont {Neupert}},
  \bibinfo {author} {\bibfnamefont {C.}~\bibnamefont {Chamon}},\ and\ \bibinfo
  {author} {\bibfnamefont {C.}~\bibnamefont {Mudry}},\ }\href
  {https://doi.org/10.1103/PhysRevB.99.245138} {\bibfield  {journal} {\bibinfo
  {journal} {Phys. Rev. B}\ }\textbf {\bibinfo {volume} {99}},\ \bibinfo
  {pages} {245138} (\bibinfo {year} {2019})}\BibitemShut {NoStop}%
\bibitem [{\citenamefont {Zhou}\ and\ \citenamefont {Guo}(2019)}]{Zhou2019}%
  \BibitemOpen
  \bibfield  {author} {\bibinfo {author} {\bibfnamefont {C.}~\bibnamefont
  {Zhou}}\ and\ \bibinfo {author} {\bibfnamefont {H.}~\bibnamefont {Guo}},\
  }\href {https://doi.org/10.1103/PhysRevB.99.035423} {\bibfield  {journal}
  {\bibinfo  {journal} {Phys. Rev. B}\ }\textbf {\bibinfo {volume} {99}},\
  \bibinfo {pages} {035423} (\bibinfo {year} {2019})}\BibitemShut {NoStop}%
\bibitem [{\citenamefont {Debray}\ \emph {et~al.}(2002)\citenamefont {Debray},
  \citenamefont {Zverev}, \citenamefont {Gurevich}, \citenamefont {Klesse},\
  and\ \citenamefont {Newrock}}]{Debray2002}%
  \BibitemOpen
  \bibfield  {author} {\bibinfo {author} {\bibfnamefont {P.}~\bibnamefont
  {Debray}}, \bibinfo {author} {\bibfnamefont {V.~N.}\ \bibnamefont {Zverev}},
  \bibinfo {author} {\bibfnamefont {V.}~\bibnamefont {Gurevich}}, \bibinfo
  {author} {\bibfnamefont {R.}~\bibnamefont {Klesse}},\ and\ \bibinfo {author}
  {\bibfnamefont {R.~S.}\ \bibnamefont {Newrock}},\ }\href
  {https://doi.org/10.1088/0268-1242/17/11/201} {\bibfield  {journal} {\bibinfo
   {journal} {Semicond. Sci. Technol.}\ }\textbf {\bibinfo {volume} {17}},\
  \bibinfo {pages} {R21} (\bibinfo {year} {2002})}\BibitemShut {NoStop}%
\bibitem [{\citenamefont {Tanatar}(1998)}]{Tanatar1998b}%
  \BibitemOpen
  \bibfield  {author} {\bibinfo {author} {\bibfnamefont {B.}~\bibnamefont
  {Tanatar}},\ }\href {https://doi.org/10.1103/PhysRevB.58.1154} {\bibfield
  {journal} {\bibinfo  {journal} {Phys. Rev. B}\ }\textbf {\bibinfo {volume}
  {58}},\ \bibinfo {pages} {1154} (\bibinfo {year} {1998})}\BibitemShut
  {NoStop}%
\bibitem [{\citenamefont {Misquitta}\ \emph {et~al.}(2014)\citenamefont
  {Misquitta}, \citenamefont {Maezono}, \citenamefont {Drummond}, \citenamefont
  {Stone},\ and\ \citenamefont {Needs}}]{Misquitta2014}%
  \BibitemOpen
  \bibfield  {author} {\bibinfo {author} {\bibfnamefont {A.~J.}\ \bibnamefont
  {Misquitta}}, \bibinfo {author} {\bibfnamefont {R.}~\bibnamefont {Maezono}},
  \bibinfo {author} {\bibfnamefont {N.~D.}\ \bibnamefont {Drummond}}, \bibinfo
  {author} {\bibfnamefont {A.~J.}\ \bibnamefont {Stone}},\ and\ \bibinfo
  {author} {\bibfnamefont {R.~J.}\ \bibnamefont {Needs}},\ }\href
  {https://doi.org/10.1103/PhysRevB.89.045140} {\bibfield  {journal} {\bibinfo
  {journal} {Phys. Rev. B}\ }\textbf {\bibinfo {volume} {89}},\ \bibinfo
  {pages} {045140} (\bibinfo {year} {2014})}\BibitemShut {NoStop}%
\bibitem [{\citenamefont {Drummond}\ and\ \citenamefont
  {Needs}(2007)}]{Drummond2007}%
  \BibitemOpen
  \bibfield  {author} {\bibinfo {author} {\bibfnamefont {N.~D.}\ \bibnamefont
  {Drummond}}\ and\ \bibinfo {author} {\bibfnamefont {R.~J.}\ \bibnamefont
  {Needs}},\ }\href {https://doi.org/10.1103/PhysRevLett.99.166401} {\bibfield
  {journal} {\bibinfo  {journal} {Phys. Rev. Lett.}\ }\textbf {\bibinfo
  {volume} {99}},\ \bibinfo {pages} {166401} (\bibinfo {year}
  {2007})}\BibitemShut {NoStop}%
\bibitem [{\citenamefont {Dobson}\ \emph {et~al.}(2006)\citenamefont {Dobson},
  \citenamefont {White},\ and\ \citenamefont {Rubio}}]{Dobson2006}%
  \BibitemOpen
  \bibfield  {author} {\bibinfo {author} {\bibfnamefont {J.~F.}\ \bibnamefont
  {Dobson}}, \bibinfo {author} {\bibfnamefont {A.}~\bibnamefont {White}},\ and\
  \bibinfo {author} {\bibfnamefont {A.}~\bibnamefont {Rubio}},\ }\href
  {https://doi.org/10.1103/PhysRevLett.96.073201} {\bibfield  {journal}
  {\bibinfo  {journal} {Phys. Rev. Lett.}\ }\textbf {\bibinfo {volume} {96}},\
  \bibinfo {pages} {073201} (\bibinfo {year} {2006})}\BibitemShut {NoStop}%
\bibitem [{\citenamefont {Chang}\ \emph {et~al.}(1971)\citenamefont {Chang},
  \citenamefont {Cooper}, \citenamefont {Drummond},\ and\ \citenamefont
  {Young}}]{Chang1971}%
  \BibitemOpen
  \bibfield  {author} {\bibinfo {author} {\bibfnamefont {D.}~\bibnamefont
  {Chang}}, \bibinfo {author} {\bibfnamefont {R.}~\bibnamefont {Cooper}},
  \bibinfo {author} {\bibfnamefont {J.}~\bibnamefont {Drummond}},\ and\
  \bibinfo {author} {\bibfnamefont {A.}~\bibnamefont {Young}},\ }\href
  {https://doi.org/10.1016/0375-9601(71)90685-2} {\bibfield  {journal}
  {\bibinfo  {journal} {Phys. Lett. A}\ }\textbf {\bibinfo {volume} {37}},\
  \bibinfo {pages} {311} (\bibinfo {year} {1971})}\BibitemShut {NoStop}%
\bibitem [{\citenamefont {Moudgil}\ \emph
  {et~al.}(2010{\natexlab{b}})\citenamefont {Moudgil}, \citenamefont {Garg},\
  and\ \citenamefont {Ahluwalia}}]{Moudgil2010}%
  \BibitemOpen
  \bibfield  {author} {\bibinfo {author} {\bibfnamefont {R.~K.}\ \bibnamefont
  {Moudgil}}, \bibinfo {author} {\bibfnamefont {V.}~\bibnamefont {Garg}},\ and\
  \bibinfo {author} {\bibfnamefont {P.~K.}\ \bibnamefont {Ahluwalia}},\ }\href
  {https://doi.org/10.1140/epjb/e2010-00103-9} {\bibfield  {journal} {\bibinfo
  {journal} {Eur. Phys. J. B}\ }\textbf {\bibinfo {volume} {74}},\ \bibinfo
  {pages} {517} (\bibinfo {year} {2010}{\natexlab{b}})}\BibitemShut {NoStop}%
\bibitem [{\citenamefont {Zhang}\ \emph {et~al.}(2008)\citenamefont {Zhang},
  \citenamefont {Shen},\ and\ \citenamefont {Liu}}]{Zhang2008}%
  \BibitemOpen
  \bibfield  {author} {\bibinfo {author} {\bibfnamefont {H.}~\bibnamefont
  {Zhang}}, \bibinfo {author} {\bibfnamefont {M.}~\bibnamefont {Shen}},\ and\
  \bibinfo {author} {\bibfnamefont {J.}~\bibnamefont {Liu}},\ }\href
  {https://doi.org/10.1063/1.2874115} {\bibfield  {journal} {\bibinfo
  {journal} {J. Appl. Phys.}\ }\textbf {\bibinfo {volume} {103}},\ \bibinfo
  {pages} {043705} (\bibinfo {year} {2008})}\BibitemShut {NoStop}%
\bibitem [{\citenamefont {Szafran}\ \emph {et~al.}(2005)\citenamefont
  {Szafran}, \citenamefont {Chwiej}, \citenamefont {Peeters}, \citenamefont
  {Bednarek},\ and\ \citenamefont {Adamowski}}]{Szafran2005}%
  \BibitemOpen
  \bibfield  {author} {\bibinfo {author} {\bibfnamefont {B.}~\bibnamefont
  {Szafran}}, \bibinfo {author} {\bibfnamefont {T.}~\bibnamefont {Chwiej}},
  \bibinfo {author} {\bibfnamefont {F.~M.}\ \bibnamefont {Peeters}}, \bibinfo
  {author} {\bibfnamefont {S.}~\bibnamefont {Bednarek}},\ and\ \bibinfo
  {author} {\bibfnamefont {J.}~\bibnamefont {Adamowski}},\ }\href
  {https://doi.org/10.1103/PhysRevB.71.235305} {\bibfield  {journal} {\bibinfo
  {journal} {Phys. Rev. B}\ }\textbf {\bibinfo {volume} {71}},\ \bibinfo
  {pages} {235305} (\bibinfo {year} {2005})}\BibitemShut {NoStop}%
\bibitem [{\citenamefont {Tsuchiya}(2001)}]{Tsuchiya2001}%
  \BibitemOpen
  \bibfield  {author} {\bibinfo {author} {\bibfnamefont {T.}~\bibnamefont
  {Tsuchiya}},\ }\href {https://doi.org/10.1142/S0217979201009165} {\bibfield
  {journal} {\bibinfo  {journal} {Int. J. Mod. Phys. B}\ }\textbf {\bibinfo
  {volume} {15}},\ \bibinfo {pages} {3985} (\bibinfo {year}
  {2001})}\BibitemShut {NoStop}%
\bibitem [{\citenamefont {Gold}(1992)}]{Gold1992}%
  \BibitemOpen
  \bibfield  {author} {\bibinfo {author} {\bibfnamefont {A.}~\bibnamefont
  {Gold}},\ }\href {https://doi.org/10.1080/09500839208219028} {\bibfield
  {journal} {\bibinfo  {journal} {Philos. Mag. Lett.}\ }\textbf {\bibinfo
  {volume} {66}},\ \bibinfo {pages} {163} (\bibinfo {year} {1992})}\BibitemShut
  {NoStop}%
\bibitem [{\citenamefont {Saini}\ \emph {et~al.}(2004)\citenamefont {Saini},
  \citenamefont {Tankeshwar},\ and\ \citenamefont {Moudgil}}]{Saini2004}%
  \BibitemOpen
  \bibfield  {author} {\bibinfo {author} {\bibfnamefont {L.~K.}\ \bibnamefont
  {Saini}}, \bibinfo {author} {\bibfnamefont {K.}~\bibnamefont {Tankeshwar}},\
  and\ \bibinfo {author} {\bibfnamefont {R.~K.}\ \bibnamefont {Moudgil}},\
  }\href {https://doi.org/10.1103/PhysRevB.70.075302} {\bibfield  {journal}
  {\bibinfo  {journal} {Phys. Rev. B}\ }\textbf {\bibinfo {volume} {70}},\
  \bibinfo {pages} {075302} (\bibinfo {year} {2004})}\BibitemShut {NoStop}%
\bibitem [{\citenamefont {Moudgil}(2000)}]{Moudgil2000}%
  \BibitemOpen
  \bibfield  {author} {\bibinfo {author} {\bibfnamefont {R.~K.}\ \bibnamefont
  {Moudgil}},\ }\href {https://doi.org/10.1088/0953-8984/12/8/319} {\bibfield
  {journal} {\bibinfo  {journal} {J. Phys.: Condens. Matter}\ }\textbf
  {\bibinfo {volume} {12}},\ \bibinfo {pages} {1781} (\bibinfo {year}
  {2000})}\BibitemShut {NoStop}%
\bibitem [{\citenamefont {Mutluay}\ and\ \citenamefont
  {Tanatar}(1997)}]{Mutluay1997a}%
  \BibitemOpen
  \bibfield  {author} {\bibinfo {author} {\bibfnamefont {N.}~\bibnamefont
  {Mutluay}}\ and\ \bibinfo {author} {\bibfnamefont {B.}~\bibnamefont
  {Tanatar}},\ }\href {https://doi.org/10.1103/PhysRevB.55.6697} {\bibfield
  {journal} {\bibinfo  {journal} {Phys. Rev. B}\ }\textbf {\bibinfo {volume}
  {55}},\ \bibinfo {pages} {6697} (\bibinfo {year} {1997})}\BibitemShut
  {NoStop}%
\bibitem [{\citenamefont {Thakur}\ and\ \citenamefont
  {Neilson}(1997)}]{Thakur1997}%
  \BibitemOpen
  \bibfield  {author} {\bibinfo {author} {\bibfnamefont {J.~S.}\ \bibnamefont
  {Thakur}}\ and\ \bibinfo {author} {\bibfnamefont {D.}~\bibnamefont
  {Neilson}},\ }\href {https://doi.org/10.1103/PhysRevB.56.4671} {\bibfield
  {journal} {\bibinfo  {journal} {Phys. Rev. B}\ }\textbf {\bibinfo {volume}
  {56}},\ \bibinfo {pages} {4671} (\bibinfo {year} {1997})}\BibitemShut
  {NoStop}%
\bibitem [{\citenamefont {Wang}\ and\ \citenamefont {Ruden}(1995)}]{Wang1995a}%
  \BibitemOpen
  \bibfield  {author} {\bibinfo {author} {\bibfnamefont {R.}~\bibnamefont
  {Wang}}\ and\ \bibinfo {author} {\bibfnamefont {P.~P.}\ \bibnamefont
  {Ruden}},\ }\href {https://doi.org/10.1103/PhysRevB.52.7826} {\bibfield
  {journal} {\bibinfo  {journal} {Phys. Rev. B}\ }\textbf {\bibinfo {volume}
  {52}},\ \bibinfo {pages} {7826} (\bibinfo {year} {1995})}\BibitemShut
  {NoStop}%
\bibitem [{\citenamefont {Hansen}\ \emph {et~al.}(1987)\citenamefont {Hansen},
  \citenamefont {Horst}, \citenamefont {Kotthaus}, \citenamefont {Merkt},
  \citenamefont {Sikorski},\ and\ \citenamefont {Ploog}}]{Hansen1987}%
  \BibitemOpen
  \bibfield  {author} {\bibinfo {author} {\bibfnamefont {W.}~\bibnamefont
  {Hansen}}, \bibinfo {author} {\bibfnamefont {M.}~\bibnamefont {Horst}},
  \bibinfo {author} {\bibfnamefont {J.~P.}\ \bibnamefont {Kotthaus}}, \bibinfo
  {author} {\bibfnamefont {U.}~\bibnamefont {Merkt}}, \bibinfo {author}
  {\bibfnamefont {C.}~\bibnamefont {Sikorski}},\ and\ \bibinfo {author}
  {\bibfnamefont {K.}~\bibnamefont {Ploog}},\ }\href
  {https://doi.org/10.1103/PhysRevLett.58.2586} {\bibfield  {journal} {\bibinfo
   {journal} {Phys. Rev. Lett.}\ }\textbf {\bibinfo {volume} {58}},\ \bibinfo
  {pages} {2586} (\bibinfo {year} {1987})}\BibitemShut {NoStop}%
\bibitem [{\citenamefont {Demel}\ \emph {et~al.}(1988)\citenamefont {Demel},
  \citenamefont {Heitmann}, \citenamefont {Grambow},\ and\ \citenamefont
  {Ploog}}]{Demel1988}%
  \BibitemOpen
  \bibfield  {author} {\bibinfo {author} {\bibfnamefont {T.}~\bibnamefont
  {Demel}}, \bibinfo {author} {\bibfnamefont {D.}~\bibnamefont {Heitmann}},
  \bibinfo {author} {\bibfnamefont {P.}~\bibnamefont {Grambow}},\ and\ \bibinfo
  {author} {\bibfnamefont {K.}~\bibnamefont {Ploog}},\ }\href
  {https://doi.org/10.1103/PhysRevB.38.12732} {\bibfield  {journal} {\bibinfo
  {journal} {Phys. Rev. B}\ }\textbf {\bibinfo {volume} {38}},\ \bibinfo
  {pages} {12732} (\bibinfo {year} {1988})}\BibitemShut {NoStop}%
\bibitem [{\citenamefont {Debray}\ \emph {et~al.}(2001)\citenamefont {Debray},
  \citenamefont {Zverev}, \citenamefont {Raichev}, \citenamefont {Klesse},
  \citenamefont {Vasilopoulos},\ and\ \citenamefont {Newrock}}]{Debray2001}%
  \BibitemOpen
  \bibfield  {author} {\bibinfo {author} {\bibfnamefont {P.}~\bibnamefont
  {Debray}}, \bibinfo {author} {\bibfnamefont {V.}~\bibnamefont {Zverev}},
  \bibinfo {author} {\bibfnamefont {O.}~\bibnamefont {Raichev}}, \bibinfo
  {author} {\bibfnamefont {R.}~\bibnamefont {Klesse}}, \bibinfo {author}
  {\bibfnamefont {P.}~\bibnamefont {Vasilopoulos}},\ and\ \bibinfo {author}
  {\bibfnamefont {R.~S.}\ \bibnamefont {Newrock}},\ }\href
  {https://doi.org/10.1088/0953-8984/13/14/312} {\bibfield  {journal} {\bibinfo
   {journal} {J. Phys. Condens. Matter}\ }\textbf {\bibinfo {volume} {13}},\
  \bibinfo {pages} {3389} (\bibinfo {year} {2001})}\BibitemShut {NoStop}%
\bibitem [{\citenamefont {Saunders}\ \emph {et~al.}(1994)\citenamefont
  {Saunders}, \citenamefont {Freyria-Fava}, \citenamefont {Dovesi},\ and\
  \citenamefont {Roetti}}]{Saunders1994}%
  \BibitemOpen
  \bibfield  {author} {\bibinfo {author} {\bibfnamefont {V.}~\bibnamefont
  {Saunders}}, \bibinfo {author} {\bibfnamefont {C.}~\bibnamefont
  {Freyria-Fava}}, \bibinfo {author} {\bibfnamefont {R.}~\bibnamefont
  {Dovesi}},\ and\ \bibinfo {author} {\bibfnamefont {C.}~\bibnamefont
  {Roetti}},\ }\href {https://doi.org/10.1016/0010-4655(94)90209-7} {\bibfield
  {journal} {\bibinfo  {journal} {Comput. Phys. Commun.}\ }\textbf {\bibinfo
  {volume} {84}},\ \bibinfo {pages} {156} (\bibinfo {year} {1994})}\BibitemShut
  {NoStop}%
\bibitem [{\citenamefont {Lieb}\ and\ \citenamefont {Mattis}(1962)}]{Lieb1962}%
  \BibitemOpen
  \bibfield  {author} {\bibinfo {author} {\bibfnamefont {E.}~\bibnamefont
  {Lieb}}\ and\ \bibinfo {author} {\bibfnamefont {D.}~\bibnamefont {Mattis}},\
  }\href {https://doi.org/10.1103/PhysRev.125.164} {\bibfield  {journal}
  {\bibinfo  {journal} {Phys. Rev.}\ }\textbf {\bibinfo {volume} {125}},\
  \bibinfo {pages} {164} (\bibinfo {year} {1962})}\BibitemShut {NoStop}%
\bibitem [{\citenamefont {Needs}\ \emph {et~al.}(2020)\citenamefont {Needs},
  \citenamefont {Towler}, \citenamefont {Drummond}, \citenamefont {{L{\'{o}}pez
  R{\'{i}}os}},\ and\ \citenamefont {Trail}}]{Needs2020}%
  \BibitemOpen
  \bibfield  {author} {\bibinfo {author} {\bibfnamefont {R.~J.}\ \bibnamefont
  {Needs}}, \bibinfo {author} {\bibfnamefont {M.~D.}\ \bibnamefont {Towler}},
  \bibinfo {author} {\bibfnamefont {N.~D.}\ \bibnamefont {Drummond}}, \bibinfo
  {author} {\bibfnamefont {P.}~\bibnamefont {{L{\'{o}}pez R{\'{i}}os}}},\ and\
  \bibinfo {author} {\bibfnamefont {J.~R.}\ \bibnamefont {Trail}},\ }\href
  {https://doi.org/10.1063/1.5144288} {\bibfield  {journal} {\bibinfo
  {journal} {J. Chem. Phys.}\ }\textbf {\bibinfo {volume} {152}},\ \bibinfo
  {pages} {154106} (\bibinfo {year} {2020})}\BibitemShut {NoStop}%
\bibitem [{\citenamefont {Foulkes}\ \emph {et~al.}(2001)\citenamefont
  {Foulkes}, \citenamefont {Mitas}, \citenamefont {Needs},\ and\ \citenamefont
  {Rajagopal}}]{Foulkes2001}%
  \BibitemOpen
  \bibfield  {author} {\bibinfo {author} {\bibfnamefont {W.~M.~C.}\
  \bibnamefont {Foulkes}}, \bibinfo {author} {\bibfnamefont {L.}~\bibnamefont
  {Mitas}}, \bibinfo {author} {\bibfnamefont {R.~J.}\ \bibnamefont {Needs}},\
  and\ \bibinfo {author} {\bibfnamefont {G.}~\bibnamefont {Rajagopal}},\ }\href
  {https://doi.org/10.1103/RevModPhys.73.33} {\bibfield  {journal} {\bibinfo
  {journal} {Rev. Mod. Phys.}\ }\textbf {\bibinfo {volume} {73}},\ \bibinfo
  {pages} {33} (\bibinfo {year} {2001})}\BibitemShut {NoStop}%
\bibitem [{\citenamefont {{L{\'{o}}pez R{\'{i}}os}}\ \emph
  {et~al.}(2006)\citenamefont {{L{\'{o}}pez R{\'{i}}os}}, \citenamefont {Ma},
  \citenamefont {Drummond}, \citenamefont {Towler},\ and\ \citenamefont
  {Needs}}]{LopezRios2006a}%
  \BibitemOpen
  \bibfield  {author} {\bibinfo {author} {\bibfnamefont {P.}~\bibnamefont
  {{L{\'{o}}pez R{\'{i}}os}}}, \bibinfo {author} {\bibfnamefont
  {A.}~\bibnamefont {Ma}}, \bibinfo {author} {\bibfnamefont {N.~D.}\
  \bibnamefont {Drummond}}, \bibinfo {author} {\bibfnamefont {M.~D.}\
  \bibnamefont {Towler}},\ and\ \bibinfo {author} {\bibfnamefont {R.~J.}\
  \bibnamefont {Needs}},\ }\href {https://doi.org/10.1103/PhysRevE.74.066701}
  {\bibfield  {journal} {\bibinfo  {journal} {Phys. Rev. E}\ }\textbf {\bibinfo
  {volume} {74}},\ \bibinfo {pages} {066701} (\bibinfo {year}
  {2006})}\BibitemShut {NoStop}%
\bibitem [{\citenamefont {Drummond}\ \emph {et~al.}(2004)\citenamefont
  {Drummond}, \citenamefont {Towler},\ and\ \citenamefont
  {Needs}}]{Drummond2004a}%
  \BibitemOpen
  \bibfield  {author} {\bibinfo {author} {\bibfnamefont {N.~D.}\ \bibnamefont
  {Drummond}}, \bibinfo {author} {\bibfnamefont {M.~D.}\ \bibnamefont
  {Towler}},\ and\ \bibinfo {author} {\bibfnamefont {R.~J.}\ \bibnamefont
  {Needs}},\ }\href {https://doi.org/10.1103/PhysRevB.70.235119} {\bibfield
  {journal} {\bibinfo  {journal} {Phys. Rev. B}\ }\textbf {\bibinfo {volume}
  {70}},\ \bibinfo {pages} {235119} (\bibinfo {year} {2004})}\BibitemShut
  {NoStop}%
\bibitem [{\citenamefont {Umrigar}\ \emph {et~al.}(1988)\citenamefont
  {Umrigar}, \citenamefont {Wilson},\ and\ \citenamefont
  {Wilkins}}]{Umrigar1988}%
  \BibitemOpen
  \bibfield  {author} {\bibinfo {author} {\bibfnamefont {C.~J.}\ \bibnamefont
  {Umrigar}}, \bibinfo {author} {\bibfnamefont {K.~G.}\ \bibnamefont
  {Wilson}},\ and\ \bibinfo {author} {\bibfnamefont {J.~W.}\ \bibnamefont
  {Wilkins}},\ }\href {https://doi.org/10.1103/PhysRevLett.60.1719} {\bibfield
  {journal} {\bibinfo  {journal} {Phys. Rev. Lett.}\ }\textbf {\bibinfo
  {volume} {60}},\ \bibinfo {pages} {1719} (\bibinfo {year}
  {1988})}\BibitemShut {NoStop}%
\bibitem [{\citenamefont {Drummond}\ and\ \citenamefont
  {Needs}(2005)}]{Drummond2005}%
  \BibitemOpen
  \bibfield  {author} {\bibinfo {author} {\bibfnamefont {N.~D.}\ \bibnamefont
  {Drummond}}\ and\ \bibinfo {author} {\bibfnamefont {R.~J.}\ \bibnamefont
  {Needs}},\ }\href {https://doi.org/10.1103/PhysRevB.72.085124} {\bibfield
  {journal} {\bibinfo  {journal} {Phys. Rev. B}\ }\textbf {\bibinfo {volume}
  {72}},\ \bibinfo {pages} {085124} (\bibinfo {year} {2005})}\BibitemShut
  {NoStop}%
\bibitem [{\citenamefont {Umrigar}\ \emph {et~al.}(2007)\citenamefont
  {Umrigar}, \citenamefont {Toulouse}, \citenamefont {Filippi}, \citenamefont
  {Sorella},\ and\ \citenamefont {Hennig}}]{Umrigar2007}%
  \BibitemOpen
  \bibfield  {author} {\bibinfo {author} {\bibfnamefont {C.~J.}\ \bibnamefont
  {Umrigar}}, \bibinfo {author} {\bibfnamefont {J.}~\bibnamefont {Toulouse}},
  \bibinfo {author} {\bibfnamefont {C.}~\bibnamefont {Filippi}}, \bibinfo
  {author} {\bibfnamefont {S.}~\bibnamefont {Sorella}},\ and\ \bibinfo {author}
  {\bibfnamefont {R.~G.}\ \bibnamefont {Hennig}},\ }\href
  {https://doi.org/10.1103/PhysRevLett.98.110201} {\bibfield  {journal}
  {\bibinfo  {journal} {Phys. Rev. Lett.}\ }\textbf {\bibinfo {volume} {98}},\
  \bibinfo {pages} {110201} (\bibinfo {year} {2007})}\BibitemShut {NoStop}%
\bibitem [{\citenamefont {Lin}\ \emph {et~al.}(2001)\citenamefont {Lin},
  \citenamefont {Zong},\ and\ \citenamefont {Ceperley}}]{Lin2001}%
  \BibitemOpen
  \bibfield  {author} {\bibinfo {author} {\bibfnamefont {C.}~\bibnamefont
  {Lin}}, \bibinfo {author} {\bibfnamefont {F.~H.}\ \bibnamefont {Zong}},\ and\
  \bibinfo {author} {\bibfnamefont {D.~M.}\ \bibnamefont {Ceperley}},\ }\href
  {https://doi.org/10.1103/PhysRevE.64.016702} {\bibfield  {journal} {\bibinfo
  {journal} {Phys. Rev. E}\ }\textbf {\bibinfo {volume} {64}},\ \bibinfo
  {pages} {016702} (\bibinfo {year} {2001})}\BibitemShut {NoStop}%
\bibitem [{\citenamefont {Shulenburger}\ \emph {et~al.}(2009)\citenamefont
  {Shulenburger}, \citenamefont {Casula}, \citenamefont {Senatore},\ and\
  \citenamefont {Martin}}]{Shulenburger2009}%
  \BibitemOpen
  \bibfield  {author} {\bibinfo {author} {\bibfnamefont {L.}~\bibnamefont
  {Shulenburger}}, \bibinfo {author} {\bibfnamefont {M.}~\bibnamefont
  {Casula}}, \bibinfo {author} {\bibfnamefont {G.}~\bibnamefont {Senatore}},\
  and\ \bibinfo {author} {\bibfnamefont {R.~M.}\ \bibnamefont {Martin}},\
  }\href {https://doi.org/10.1088/1751-8113/42/21/214021} {\bibfield  {journal}
  {\bibinfo  {journal} {J. Phys. A Math. Theor.}\ }\textbf {\bibinfo {volume}
  {42}},\ \bibinfo {pages} {214021} (\bibinfo {year} {2009})}\BibitemShut
  {NoStop}%
\bibitem [{fit()}]{fit}%
  \BibitemOpen
  \href@noop {} {}\bibinfo {note} {GNUPLOT and Mathematica softwares are used
  for fittings.}\BibitemShut {Stop}%
\bibitem [{\citenamefont {Fabbri}\ \emph {et~al.}(2015)\citenamefont {Fabbri},
  \citenamefont {Panfil}, \citenamefont {Cl{\'{e}}ment}, \citenamefont
  {Fallani}, \citenamefont {Inguscio}, \citenamefont {Fort},\ and\
  \citenamefont {Caux}}]{Fabbri2015}%
  \BibitemOpen
  \bibfield  {author} {\bibinfo {author} {\bibfnamefont {N.}~\bibnamefont
  {Fabbri}}, \bibinfo {author} {\bibfnamefont {M.}~\bibnamefont {Panfil}},
  \bibinfo {author} {\bibfnamefont {D.}~\bibnamefont {Cl{\'{e}}ment}}, \bibinfo
  {author} {\bibfnamefont {L.}~\bibnamefont {Fallani}}, \bibinfo {author}
  {\bibfnamefont {M.}~\bibnamefont {Inguscio}}, \bibinfo {author}
  {\bibfnamefont {C.}~\bibnamefont {Fort}},\ and\ \bibinfo {author}
  {\bibfnamefont {J.-S.}\ \bibnamefont {Caux}},\ }\href
  {https://doi.org/10.1103/PhysRevA.91.043617} {\bibfield  {journal} {\bibinfo
  {journal} {Phys. Rev. A}\ }\textbf {\bibinfo {volume} {91}},\ \bibinfo
  {pages} {043617} (\bibinfo {year} {2015})},\ \Eprint
  {https://arxiv.org/abs/1406.2176} {1406.2176} \BibitemShut {NoStop}%
\bibitem [{\citenamefont {Mattis}\ and\ \citenamefont
  {Lieb}(1965)}]{Mattis1965}%
  \BibitemOpen
  \bibfield  {author} {\bibinfo {author} {\bibfnamefont {D.~C.}\ \bibnamefont
  {Mattis}}\ and\ \bibinfo {author} {\bibfnamefont {E.~H.}\ \bibnamefont
  {Lieb}},\ }\href {https://doi.org/10.1063/1.1704281} {\bibfield  {journal}
  {\bibinfo  {journal} {J. Math. Phys.}\ }\textbf {\bibinfo {volume} {6}},\
  \bibinfo {pages} {304} (\bibinfo {year} {1965})}\BibitemShut {NoStop}%
\bibitem [{\citenamefont {Schulz}(1990)}]{Schulz1990}%
  \BibitemOpen
  \bibfield  {author} {\bibinfo {author} {\bibfnamefont {H.~J.}\ \bibnamefont
  {Schulz}},\ }\href {https://doi.org/10.1103/PhysRevLett.64.2831} {\bibfield
  {journal} {\bibinfo  {journal} {Phys. Rev. Lett.}\ }\textbf {\bibinfo
  {volume} {64}},\ \bibinfo {pages} {2831} (\bibinfo {year}
  {1990})}\BibitemShut {NoStop}%
\end{thebibliography}%
\bibliographystyle{apsrev4-2}
\end{document}